\RequirePackage{fix-cm}
\documentclass[twocolumn,epjc3]{svjour3}

\smartqed  
\usepackage[T1]{fontenc}
\usepackage[utf8]{inputenc}
\usepackage{graphicx}
\usepackage{amsmath}
\usepackage{amssymb}
\usepackage{hyperref}
\usepackage{cite}
\usepackage{xcolor}
\usepackage[UKenglish]{babel}
\usepackage{booktabs}
\usepackage{subcaption} 
\usepackage{multirow}
\usepackage{orcidlink}

\journalname{Eur. Phys. J. C}

\begin{document}

\title{Lund Plane to Bloch (LP2B) Encoding for Object and Polarization Tagging with Quantum Jet Substructure} 


\author{Fabrizio Napolitano\thanksref{e1,addr1,addr2}\orcidlink{0000-0002-8686-5923}
        \and
        Luca Della Penna\thanksref{e2,addr1,addr2}\orcidlink{0009-0007-8978-1067}
        \and
        Tommaso Tedeschi\thanksref{e3,addr2}\orcidlink{0000-0002-7125-2905}
        \and 
        Livio Fanò\thanksref{addr1,addr2}\orcidlink{0000-0002-9007-629X}
}

\thankstext{e1}{e-mail: fabrizio.napolitano@unipg.it}
\thankstext{e2}{e-mail: luca.dellapenna@pg.infn.it}
\thankstext{e3}{e-mail: tommaso.tedeschi@pg.infn.it}

\institute{Università degli Studi di Perugia, Dipartimento di Fisica e Geologia, Via A. Pascoli, 06123, Perugia (PG), Italy \label{addr1}
\and
INFN Sezione di Perugia, Via A. Pascoli, 06123 Perugia (PG), Italy \label{addr2}
}

\date{Received: date / Accepted: date}

\maketitle

\begin{abstract}

The application of quantum algorithms to jet substructure analysis is of growing interest as Noisy Intermediate-Scale Quantum (NISQ) hardware continues to mature in both qubit count and gate depth. Jet substructure remains essential for addressing demanding and complementary challenges at the Large Hadron Collider (LHC) and beyond, notably object classification and polarization tagging. However, existing quantum machine learning approaches typically rely on data representations that suffer from infrared and collinear (IRC) unsafety, sensitivity to non-perturbative effects, or poor scalability. In this work, we introduce the Lund Plane to Bloch (LP2B) encoding, designed to map a theoretically clean and robust representation of jet kinematics directly into qubit states. Leveraging this encoding, we implement a Quantum Tree-Topology Network (QTTN) that natively embeds the hierarchical structure of the Lund tree.
We evaluate the QTTN across multiple benchmarks and observe that it matches the performance of large classical deep learning architectures, such as LundNet, on polarization tagging, while maintaining competitive accuracy for $W$ boson and top quark tagging. The architecture demonstrates enhanced sensitivity compared to standard ``one particle - one qubit'' (1P1Q) encodings on both polarization and $W$ tagging, and pushes the performance-to-cost Pareto front when compared against multi-layer perceptrons of similar size and boosted decision trees. Remarkably, the QTTN requires three orders of magnitude fewer parameters than LundNet, demonstrating promises for low-latency FPGA implementations in trigger systems. Furthermore, the QTTN outperforms classical methods in the low-data regime, making it highly suitable for low-yield, data-driven analyses. We also find that the quantum model is less susceptible to overfitting generator-specific parton shower and hadronization models than classical deep learning approaches, pointing toward potentially smaller systematic uncertainties. Finally, we successfully validate the QTTN on real quantum hardware using a 3-qubit solid-state NMR SpinQ device.
\end{abstract}

\section{Introduction}
\label{sec:intro}

The study of jet substructure has become a cornerstone of the physics program at the LHC. As collision energies and instantaneous luminosities increase, the ability to accurately identify heavily boosted decaying resonances, such as $W/Z$ bosons and top quarks, within dense QCD environments is crucial. Beyond standard object classification, jet substructure provides access to subtle kinematic correlations, enabling complex tasks such as polarization tagging. For instance, in Vector Boson Scattering (VBS) topologies, measuring the scattering amplitude of longitudinal polarized bosons ($W_L W_L \rightarrow W_L W_L$) is essential for probing the electroweak symmetry breaking  mechanism; any deviation from Standard Model (SM) couplings would lead to unitarity violations at high energies, making VBS a highly sensitive probe for Beyond the Standard Model (BSM) physics \cite{Chang:2013aya,Kilian:2014zja,PhysRevD.100.096003}. However, extracting such signals is notoriously difficult, as the final states are dominated by massive combinatorial backgrounds and complex multi-jet environments governed by QCD radiation.

Analyzing these complex hadronic signatures increasingly relies on advanced machine learning (ML) techniques. Classical algorithms, ranging from Boosted Decision Trees (BDTs) to deep learning architectures like Graph Neural Networks (GNNs) \cite{Qu:2020aa} and Transformer models \cite{Qu:2022aa}, have significantly advanced jet tagging performance. Methods based on the Lund Jet Plane \cite{dreyer2018lund,dreyer2021jet} have proven effective, as they capture the QCD clustering history within a theoretical framework that respects the physical symmetries of the radiation cascade. 

Despite their success, classical deep learning approaches in this domain face several critical challenges. First, highly parameterized models are prone to learning the production dynamics of a specific physical process, such as the distinctive transverse momentum ($p_T$) spectrum of a resonance, rather than the fundamental physics of the decay kinematics. Second, they frequently exhibit a strong dependence on generator-specific non-perturbative effects, such as parton shower and hadronization tunings, leading to elevated systematic uncertainties when applied in analyses. Finally, the massive parameter count of state-of-the-art architectures poses significant latency and scalability issues for deployment in low-level trigger environments.

In recent years, quantum machine learning (QML) is emerging as a promising tool for high-energy physics data analysis, motivated by the rapid scaling of Noisy Intermediate-Scale Quantum (NISQ) hardware in both qubit count and gate fidelity. A variety of quantum architectures have been proposed for particle classification, notably direct constituent embeddings like the ``one particle - one qubit'' (1P1Q) approach \cite{bal2025one}. While 1P1Q provides an intuitive mapping of collision events, current hardware constraints force the truncation of the jet constituents. This severely limits the model's capacity to capture the full complexity of jet substructure and leaves unaddressed the issue of Infrared and Collinear (IRC) safety, which is required for robust theoretical predictions.

We address these challenges by introducing a fundamentally different quantum representation. Rather than encoding bare particle kinematics, we base our representation on the declustering history of the jet, mapped onto the Lund Jet Plane~\cite{dreyer2018lund}. By utilizing a binary tree constructed from the Cambridge/Aachen (C/A) sequential recombination algorithm, the formulation is inherently robust against soft emissions and collinear splittings. To process this IRC-safe graph, we construct a Quantum Tree-Topology Network (QTTN), where the hierarchical topology of the QTTN naturally mirrors the physical QCD splitting cascade, allowing quantum entanglement to propagate from the leaves (the latest splittings) to the root of the tree (the initial hard parton). 

To embed this structure, we introduce the Lund Plane to Bloch (LP2B) encoding. This method utilizes a continuous, differentially deformable stereographic projection that maps the planar Lund coordinates directly onto the Bloch sphere. Guided by learnable stretch parameters, this encoding dynamically optimizes the classical phase space mapping during training, smoothly handling kinematic dead-ends without breaking the network symmetry. To ensure our model learns a truly universal representation of the underlying physics, we explicitly decouple production from decay by evaluating the architecture across distinct production scenarios of $W$/Top tagging and polarization tagging (using the tree level W boson production and a BSM heavy spin-2 graviton resonance) filtered to a flat $p_T$ spectrum, without applying any reweighting.

This abstraction guarantees physical robustness while satisfying stringent present-day quantum hardware constraints, requiring a low qubit count (e.g., $N=7$ qubits for a depth-3 tree), minimal circuit depth and parameter counts of the order $10^1-10^2$. In this work, we demonstrate that:
\begin{enumerate}
    \item[(i)] the QTTN matches the performance of large classical deep learning models (such as LundNet) on polarization tagging, while maintaining highly competitive accuracy for $W$ boson and top quark classification.
    \item[(ii)] the proposed architecture consistently outperforms the standard 1P1Q embedding on both polarization and $W$ tagging, and advances the performance-to-cost Pareto front against BDTs and classical Multi-Layer Perceptrons (MLPs) of similar size.
    \item[(iii)] the LP2B encoding achieves state-of-the-art performance with three orders of magnitude fewer parameters than LundNet, demonstrating potential for low-latency FPGA implementations in hardware triggers.
    \item[(iv)] the QTTN outperforms classical methods in the low-data regime, paving the way for dedicated applications in low-yield, data-driven analyses.
    \item[(v)] the quantum model exhibits reduced susceptibility to overfitting generator-specific hadronization and parton shower models compared to classical deep learning methods, pointing toward smaller systematic uncertainties in experimental measurements.
    \item[(vi)] VBS polarization tagging can be addressed with classical and quantum ML approaches with jet substructure.
    \item[(vii)] the QTTN is fully compatible with current NISQ capabilities and has been successfully validated on real quantum hardware using a 3-qubit solid-state NMR SpinQ device.
\end{enumerate}

The paper is structured as follows. In Section \ref{sec:data}, we describe the simulated datasets for object and polarization tagging, alongside the extraction of the Lund Jet Plane tree structures. Section \ref{sec:methods} details the LP2B encoding and the QTTN architecture, and introduces the classical BDT, MLP, and LundNet baselines. Training procedures, performance metrics, and systematic comparisons are presented in Section~\ref{sec:training} and \ref{sec:results}, including the hardware validation on the SpinQ device. Finally, concluding remarks are given in Section \ref{sec:conclusions}.

\section{Data Simulation and Representation}
\label{sec:data}

To study the performance of the proposed QTTN with the LP2B encoding, we utilize the JetGame~\cite{stefano_carrazza_2019_2602515} open dataset for $W$/Top tagging, and we generate samples containing hadronically decaying $W$ bosons from different scenarios. To ensure the models learn internal jet kinematics rather than production-specific features (acting as a `process tagger'), the polarized samples were explicitly decoupled by filtering to a flat $p_T$ spectrum, without applying any reweighting. We additionally utilize two separate, physically distinct datasets to decouple the production from the decay and demonstrate the effectiveness of our approach. To assess the impact of the hadronization and parton showering systematics on the proposed method and benchmarks, we additionally produce sample with different generators. 

\subsection{$W$/Top Tagging Dataset}
The JetGame dataset contains three types of processes, simulated with  \texttt{Pythia8} (v8.223): top jets from $t\bar{t}$ events with top quarks decaying to hadrons, $W$ jets using $WW$ events with $W$ bosons decaying to hadrons, and QCD jets using dijet events. The jets are clustered with the anti-$k_t$ algorithm with R = 0.8, taking the two leading $p_T$ and central ($|y|<2.5$) ones and transverse momentum above 500 GeV. We utilized the full statistics, of about 500k jets for training, 50k jets for validation and 50k jets for test.

\subsection{Event Generation: VBS and Graviton Scenarios}
\label{subsec:generation}
\begin{sloppypar}
Parton-level events were generated using \texttt{MadGraph5\_aMC@NLO} (v3.7.0) \cite{alwall2011madgraph} at a center-of-mass energy of $\sqrt{s} = 13.6$ TeV, using the \texttt{NNPDF3.1} LO parton distribution function set \cite{ball2017nnpdf}, provided by the LHAPDF-6 package \cite{Buckley_2015}. 
\end{sloppypar}
For the polarization scenarios, separate samples were generated where the helicities of the  produced $W$ bosons were enforced to be either purely longitudinal (LL) or purely transverse (TT), implementing the decay  using \texttt{MadSpin}, that accounts for full spin-correlations and off-shell effects \cite{artoisenet2013automatic}. 
In every case, both the $W$ bosons are forced to decay hadronically ($W \to q\bar{q}$).
Three different samples are produced:
\begin{enumerate}
    \item $\mathbf{pp\rightarrow W^+W^-}$: as a Standard Model process we consider the tree level W boson production, in both the pure transverse and longitudinally polarized configuration.
    \item \textbf{Bulk RS Graviton Resonance:} a BSM scenario featuring a heavy spin-2 graviton resonance ($pp \to G_{\text{bulk}} \to W^+ W^-$). We utilize the Bulk Randall-Sundrum (RS) model parameterized by the dimensionless coupling $\tilde{k} = k/\bar{M}_{\text{Pl}}$. Because the Standard Model fields propagate in the extra-dimensional bulk, the overlap of their wavefunctions dictates that the $G_{\text{bulk}}$ resonance couples preferentially to heavy states, yielding a naturally high fraction of longitudinally polarized $W_L W_L$ pairs.
    \item \textbf{Vector Boson Scattering (VBS):} the purely electroweak $O(\alpha^4)$ same-sign $W$-boson scattering process ($pp \to W^{\pm}W^{\pm}jj$). VBS samples are used to test the model trained on the previous two datasets, and not used in the training of the models. To enhance the signal yield in the so-called boosted region, where the W bosons are reconstructed in a single ``fat'' jet,  VBS samples are produced by requiring a minimum transverse momentum of the $W$ equal to $200$ GeV.
\end{enumerate}
\begin{sloppypar}
The parton-level events were passed to \texttt{Pythia8} (v8.316) \cite{sjostrand2008brief} for parton showering, hadronization, and underlying event simulation. The generated events subsequently underwent a fast parametric detector simulation using \texttt{Delphes} (v3.5.1) \cite{de2014delphes}, configured with the \emph{delphes\_card\_CMS\_JetClassII\_onlyFatJet.tcl } card,  modeled after a general-purpose LHC detector operating in the Run 3 era. To simulate a realistic high-luminosity environment, inelastic proton-proton collisions were overlaid to model pileup. The reconstruction of the final-state hadronic objects is based on particle-flow candidates. To mitigate the substantial impact of the simulated pileup, the Pile-Up Per Particle Identification (PUPPI) algorithm \cite{Bertolini2014} is applied. 
From this pileup-subtracted set of constituents, large-R jets are clustered via the anti-$k_t$ algorithm with R = 0.8 to capture the fully hadronic decay of the boosted $W$ bosons.  To isolate valid $W$ boson candidates, a strict mass window cut is applied, requiring the jet mass to satisfy $60$~GeV $\leq m_J \leq 100$~GeV. 

\end{sloppypar}
In order to execute the decoupling strategy, the reconstructed $W$ candidates   in both the $W^+W^-$ and graviton datasets are individually filtered to ensure a flat transverse momentum spectrum in the boosted regime of $250$~GeV $\leq p_T \leq 800$~GeV and $250$~GeV $\leq p_T \leq 500$ GeV, respectively, without applying any reweighting. Evaluating the models on these flat-$p_T$ samples ensures that the discriminatory power originates exclusively from the intra-jet radiation patterns rather than the underlying cross-section kinematics of the parent processes. 
\subsection{Different Hadronization and Parton Showering models}
Alternative samples are produced with the same hard process and detector simulation, but a different hadronization and parton shower model, using \texttt{Herwig 7.3} in place of \texttt{Pythia8}, which is the baseline generator for LHC experimental simulations.
This comparison allows estimating the theoretical systematic uncertainties related to the non-perturbative regime of QCD. While the standard \texttt{Pythia} model relies on the Lund string model for hadronization and a $p_T$ ordered approach for parton shower, alternative showering models, beside the standard \texttt{Pythia8} one, are the angular-ordered shower and the dipole shower, both implemented with \texttt{Herwig} \cite{bellm2025physicsherwig7}.
By comparing different independent, physically motivated models, we can quantify the sensitivity of the results to the underlying assumptions of color coherence and fragmenting partons.

\subsection{The Lund Jet Plane Tree Representation}
\label{subsec:lundplane}
To interface the classical collision data with the quantum algorithm, the selected large-R jets are transformed into an IRC safe topological representation using the ATLAS Lund Jet Plane \cite{aad2020measurement}. 
The constituents are declustered using the Cambridge/Aachen (C/A) algorithm \cite{Dokshitzer1997}, as implemented in \texttt{FastJet} \cite{cacciari2012fastjet}. 
Following the C/A clustering, the leading jet is iteratively declustered. At each declustering step, a parent pseudojet $p$ is split into two subjets $p_1$ and $p_2$, ordered such that $p_{T,1} < p_{T,2}$. For each splitting, two kinematic observables defining the Lund Jet Plane are extracted:
\begin{align}
    x_1 &= \ln(R_0 / \Delta R) \\
    x_2 &= \ln(1 / z) 
\end{align}
where $\Delta R$ is the Euclidean distance in the rapidity-azimuth plane between $p_1$ and $p_2$, $R_0 = 1.0$ is a normalization constant, and $z = p_{T,2}/(p_{T,1} + p_{T,2})$ represents the momentum sharing fraction. We note that momentum sharing contains most of the discriminating power for polarization tagging, as opposed to other parametrizations of the Lund Jet Plane (e.g., $k_t = p_{T,2} \Delta R_{12}$) which are more relevant for e.g. quark/gluon discrimination.

This declustering process naturally yields a binary tree. To maintain a compact and uniform structure suitable for a NISQ-era quantum circuit, we truncate the declustering history at a maximum depth of $D=3$, resulting in a fully connected binary tree with $N = 2^D - 1 = 7$ nodes. 

\begin{figure}[htpb]
    \centering
    \includegraphics[width=0.55\textwidth, trim=110 80 50 50, clip]{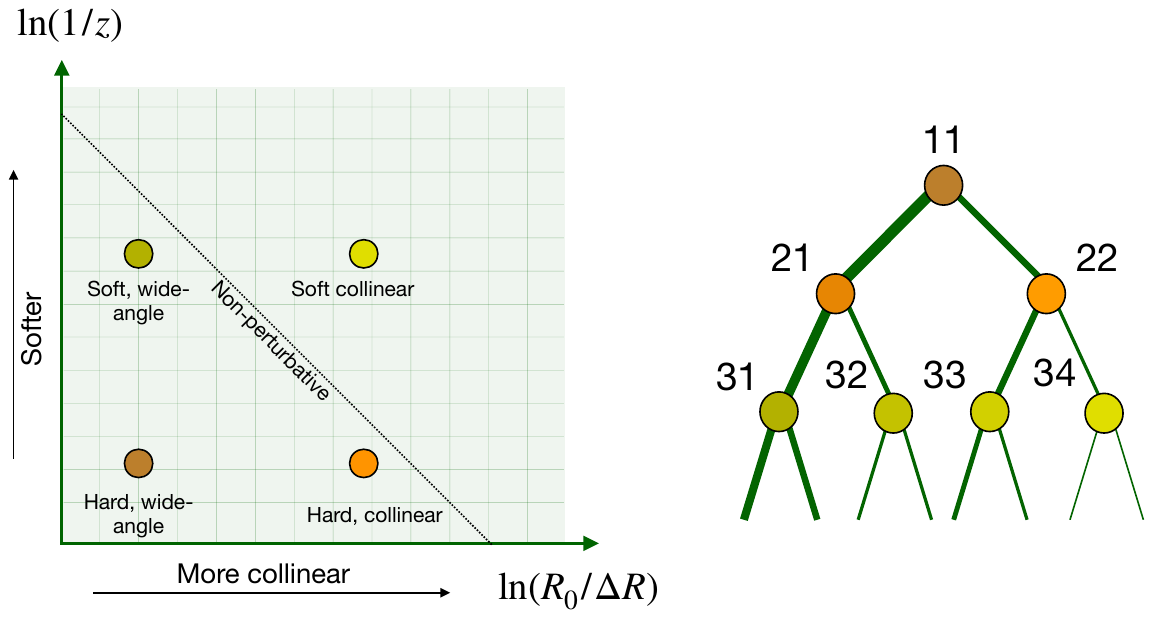}
    \caption{A visualization of the Lund Jet Plane representation and the associated binary tree. The left panel shows the Lund Jet Plane, where each point corresponds to a splitting in the jet's declustering history, in the $(\ln(R_0 / \Delta R), \ln(1 / z))$ space. Increasing values of $\ln(R_0 / \Delta R)$ correspond to more collinear splittings, while increasing $\ln(1 / z)$ indicates more asymmetric momentum sharing. The right panel illustrates the corresponding binary tree structure derived from the C/A declustering, where each node is annotated with its splitting. Empty nodes are marked with $(0.0, 0.0)$, preserving the sparsity of the jet cascade.}
    \label{fig:display_lund}
\end{figure}

\begin{figure*}[htpb]
    \centering
    \begin{subfigure}[b]{0.47\textwidth}
        \centering
        \includegraphics[width=\textwidth]{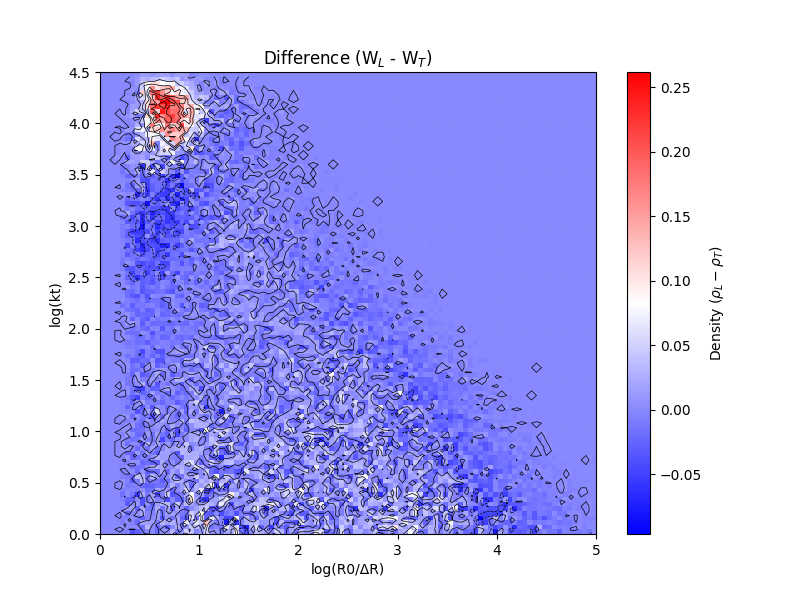}
        \caption{Density difference in the Lund Plane $\ln(k_t)$ vs $\ln(R_0/\Delta R)$}
        \label{fig:first}
    \end{subfigure}
    \hfill 
    \begin{subfigure}[b]{0.47\textwidth}
        \centering
        \includegraphics[width=\textwidth]{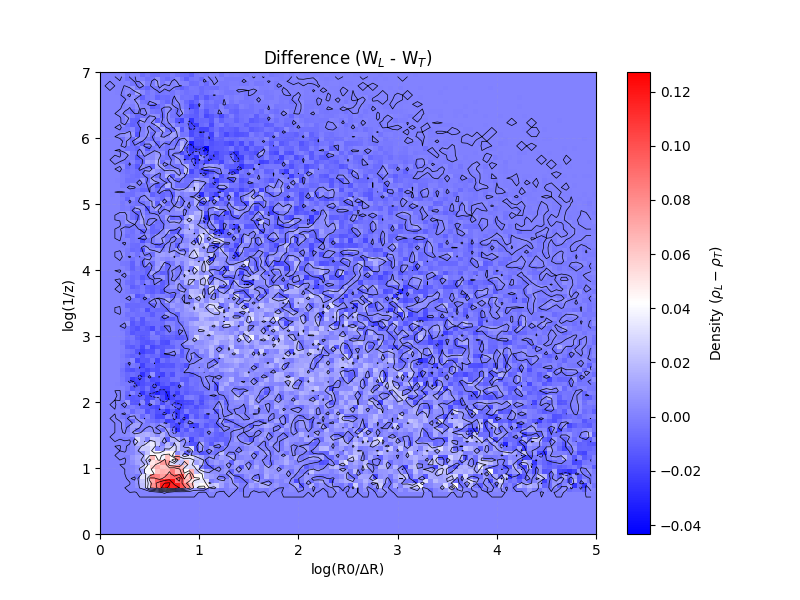}
        \caption{Density difference in the Lund Plane $\ln(1/z)$ vs $\ln(R_0/\Delta R)$}
        \label{fig:second}
    \end{subfigure}
    
    \caption{Difference in the Lund Plane (with two different parametrizations) of the densities ($\rho_L - \rho_T$) between longitudinally polarized $W_L$ jets and transversely polarized $W_T$ jets. Red regions indicate phase space where $W_L$ emissions are more frequent (typically harder and at wider angle), while blue regions indicate an excess of $W_T$ emissions (typically softer and more collinear). The QTTN exploits these distinct radiation patterns embedded in the tree structure to discriminate objects and polarization states.}
    \label{fig:lund_comparison}
\end{figure*}

A visualization of the Lund Jet Plane and the corresponding binary tree structure is provided in Figure \ref{fig:display_lund}. Each node in the tree is annotated with its corresponding Lund coordinates $(x_1, x_2)$. For transversely polarized $W$ boson, the first splitting typically happens with a softer emission and a wide angle; conversely for longitudinally polarized $W$ bosons, the first splitting is more likely to be symmetric. Figure \ref{fig:lund_comparison} shows the averaged Lund Jet Plane densities for the two polarization states, highlighting the distinct radiation patterns.

Splittings that do not occur (kinematic dead-ends or branches that terminate before depth $D=3$) are explicitly mapped to coordinates $(0.0, 0.0)$. This ensures that the topological sparsity of the QCD cascade is precisely preserved and mapped reliably into the quantum state without artificial distortions.

We apply a $\ln(k_t)$ cut to populate the Lund tree unless specified otherwise. This means that soft radiation splittings are trimmed from the tree, and the hardest branching moves upwards in the binary tree representation. As studied in~\cite{dreyer2018lund}, different values of the cut correspond to different levels of performance, and robustness on non-perturbative effects. We use $\ln(k_t)>1$ as a conservative choice, allowing keeping the tree depth contained and populated with information rich splittings.

\section{Methodology}
\label{sec:methods}

To discriminate different objects and polarization states, we process the extracted Lund Jet Plane trees using a Quantum Tree-Topology Network. 

Tree networks, and particularly tree-tensor topologies, naturally mirror the hierarchical splitting processes of the QCD parton shower, making them highly expressive for jet substructure tasks. We design the quantum circuit architecture to follow the physical C/A declustering graph, imposing a strong topological prior that minimizes the required gate count and ensures the network remains feasible for current NISQ devices.

\subsection{Quantum Tree-Topology Network}
\label{subsec:qttn}

The proposed QTTN operates on a $N=7$ qubit register, where each qubit uniquely corresponds to a node in the $D=3$ binary declustering tree. Choosing $N=7$ and $D=3$ ensure feasibility on NISQ quantum hardware without loss of generality.
The quantum model consists of an initial data encoding strategy followed by $L$ repeated variational layers. Each layer is composed of an entanglement block and a local rotation block. 

\subsubsection{Lund Plane to Bloch (LP2B) Differentiable Stereographic Encoding}
\label{subsubsec:encoding}

Mapping continuous classical variables into a bounded quantum state space (the Bloch sphere) represents a challenge in quantum machine learning. Standard angle embedding techniques often suffer from periodicity artifacts and fail to effectively handle variable-length or sparse data. To resolve this, we introduce a continuous, differentiable stereographic projection to map the Lund coordinates $(x_1, x_2)$ of each node onto the Bloch sphere.

For a given node, the classical features are mapped to two quantum rotation angles, $\theta$ and $\phi$, as follows:
\begin{align}
    r &= \sqrt{x_1^2 + x_2^2 + \epsilon} \label{eq:radius} \\
    \theta &= 2 \arctan(\lambda_i \cdot r) \label{eq:theta} \\
    \phi &= \omega_i \cdot \operatorname{atan2}(x_2, x_1) \label{eq:phi}
\end{align}
where $\epsilon = 10^{-10}$ is a small constant ensuring numerical stability during gradient backpropagation, $\lambda_i$ and $\omega_i$ are the learnable stretch parameters, specialized for each node $i$. The quantum state of each node is then initialized via the operations $R_y(\theta) R_z(\phi) |0\rangle$. 

This embedding strategy has two physical advantages. On one side, the learnable parameters  $\lambda_i$ and $\omega_i$ dynamically scale the classical Lund Jet Plane during training, specializing hierarchically for each step of the declustering history and optimally distributing the densely populated QCD phase space across the surface of the Bloch sphere without classical preprocessing. On the other, the projection preserves the structural sparsity of the jet cascade. Kinematic dead-ends (empty nodes padded with $x_1=0, x_2=0$)  yield $r=0$ and therefore $\theta=0$. Consequently, the quantum operation collapses to the identity $R_y(0)R_z(0)|0\rangle = |0\rangle$. This ``zero-safe'' behavior makes sure that absent branchings do not inject spurious rotational noise into the quantum state, preserving the IRC safety of the representation. 

\begin{figure}[htpb]
    \centering
    \includegraphics[width=0.45\textwidth, trim=50 80 0 50, clip]{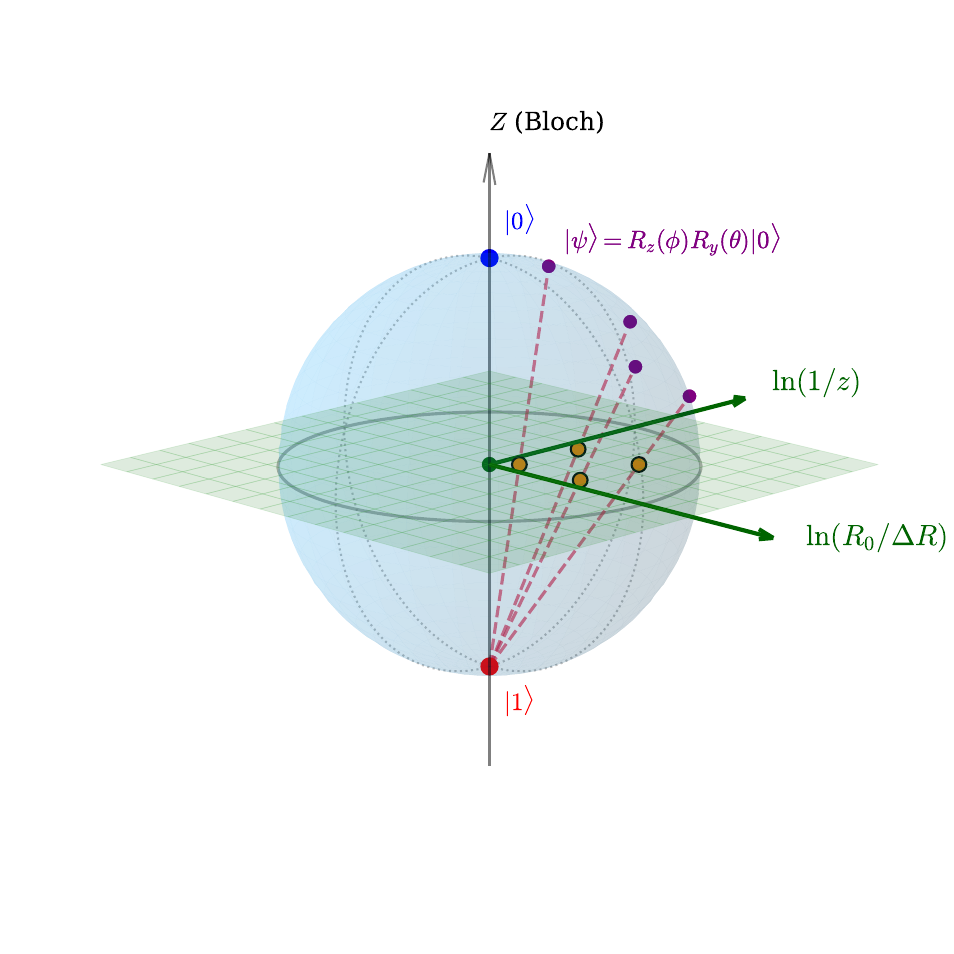}
    \caption{Geometric visualization of the differentiable stereographic projection mapping the classical Lund Jet Plane coordinates $(\ln(R_0 / \Delta R), \ln(1 / z))$ into the quantum state space. By anchoring the projection origin at the South Pole ($|1\rangle$), the mapping intrinsically handles kinematic sparsity: empty tree nodes, which carry coordinates $(0,0)$, are mapped to the North Pole ($|0\rangle$). The learnable parametes $\boldsymbol{{\lambda}}$ and $\boldsymbol{\omega}$ dynamically deform the projection during training, optimizing the coverage of the Bloch sphere.}
    \label{fig:stereographic_projection}
\end{figure}

A visualization of the LP2B encoding is shown in Figure~\ref{fig:stereographic_projection}.

\subsubsection{Variational Entanglement and Node Updates}
\label{subsubsec:variational_ansatz}

Following the state preparation, the network undergoes $L$ layers of parameterized quantum operations designed to entangle the nodes and extract hierarchical correlations.

The entanglement topology is constrained by the physical edges of the C/A declustering tree. To allow information to flow from the leaves (the latest splittings in the cascade) up to the root (the initial hard parton), we apply parameterized controlled Y rotations, $\text{CRY}(w) = \text{CNOT}\cdot\text{RY}(-w/2)\cdot\text{CNOT}\cdot\text{RY}(w/2) = \text{diag}(1,1,e^{-iw/2},e^{iw/2})$~\cite{Gheorghiu:2021aa}, along the directed edges of the tree. 
The controlled RY operation has been found to outperform simpler CNOT gates in this context. For the $N=7$ node structure, the target-control qubit pairs $(u, v)$ correspond to the physical parent-child connections: $(31, 21), (32, 21), (33, 22), (34, 22), (21, 11), (22, 11)$ as shown in Figure~\ref{fig:display_lund}, where root node $11$ is mapped to qubit 0; $21,22$ to qubit 1-2 for the first branching and $31,32,33,34$ to qubit 3-6. By restricting the entanglement topology to the C/A declustering tree rather than utilizing all-to-all connectivity, the parameter space is reduced, mitigating barren plateaus.

After the entanglement and CRY block, parametrized single-qubit rotations are applied to every qubit to explore the local Hilbert space. We utilize the general rotation gate $$Rot(\alpha, \beta, \gamma) = R_z(\alpha)R_y(\beta)R_z(\gamma)$$ where $\alpha, \beta$, and $\gamma$ are independent learnable parameters for each qubit and each of the $L-1$ layer, in case of the last layer, a $R_z(\alpha)R_y(\beta)R_x(\gamma)$ rotation is applied to the read-out qubit only.

\begin{figure*}[htpb]
    \centering
    \includegraphics[width=0.95\textwidth]{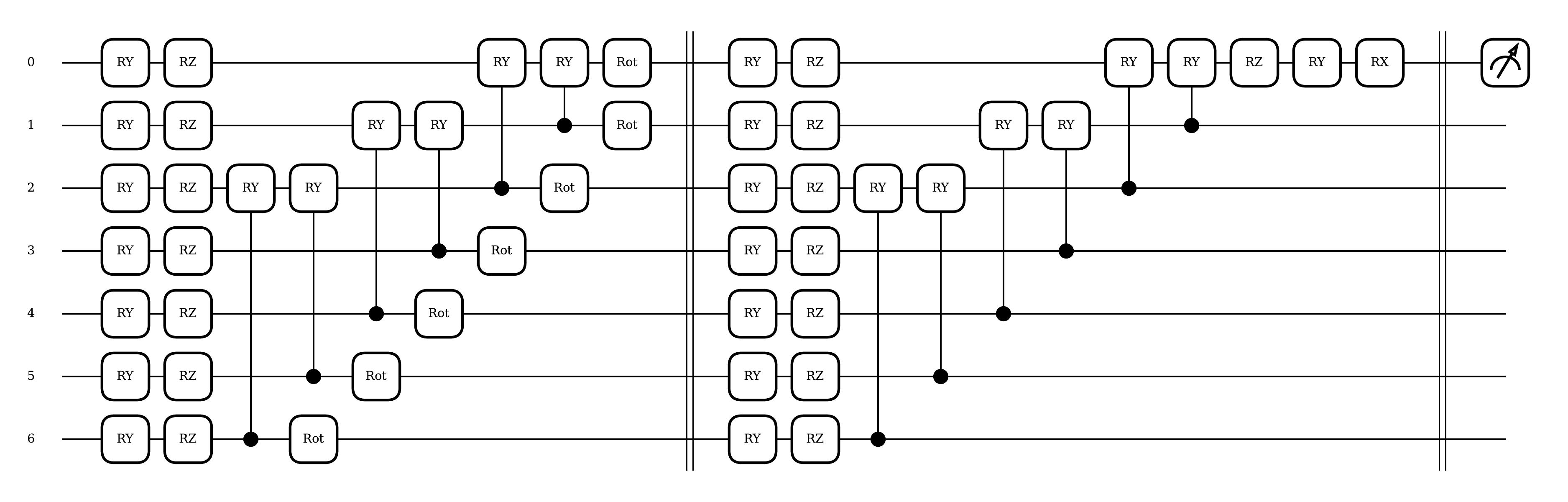}
    \caption{Schematic representation of the QTTN executed with two layers ($L=2$) on $N=7$ qubits. The circuit is divided into three functional blocks: data encoding, where the stereographically mapped Lund coordinates $(\theta, \phi)$ are loaded via $R_y$ and $R_z$ gates;  tree entanglement, where parameterized CRY gates entangle qubits following the physical edges of the C/A declustering tree; and local node updates, which apply parametrized $Rot$ gates to explore the local Hilbert space. The full architecture repeats this variational sequence for $L-1$ layers. At the last layer, rotations are applied only to the read-out qubit ($q_0$) before measuring the Pauli-Z expectation value. Qubits 0–6 correspond to Lund nodes 11, 21, 22, 31, 32, 33, and 34 respectively.}
    \label{fig:quantum_circuit}
\end{figure*}

A schematic overview of the QTTN with $L=2$ is shown in Figure~\ref{fig:quantum_circuit}.

\subsubsection{Measurement and Classical Post-Processing}
\label{subsubsec:measurement}

Since the QTTN routes information hierarchically toward the top of the cascade, the latent representation of the entire jet structure is by construction localized at the root node. Therefore, to evaluate the network, we measure the expectation value of the Pauli-Z observable exclusively on the root qubit ($q_0$):
\begin{equation}
    E_0(\boldsymbol{\Theta}) = \langle \psi(\boldsymbol{x}, \boldsymbol{\Theta}) | Z_0 | \psi(\boldsymbol{x}, \boldsymbol{\Theta}) \rangle
\end{equation}
where $|\psi(\boldsymbol{x}, \boldsymbol{\Theta})\rangle$ is the final quantum state parameterized by the classical input $\boldsymbol{x}$ and all trainable circuit weights $\boldsymbol{\Theta} = \{\boldsymbol{\lambda}, \boldsymbol{\omega}, \boldsymbol{w}, \boldsymbol{\alpha}, \boldsymbol{\beta}, \boldsymbol{\gamma}\}$. 

Finally, the quantum expectation value $E_0 \in [-1, 1]$ is shifted and scaled via a classical linear layer to produce the final continuous logit:
\begin{equation}
    y_{\text{logit}} = c_w E_0 + c_b
\end{equation}
where $c_w$ and $c_b$ are a learnable scalar weight and bias, respectively. This logit is subsequently passed through a standard sigmoid activation function to compute the binary cross-entropy loss utilized during the optimization process.

We cross-check that the quantum circuit in the QTTN formulation is expressive, and that the gates are not redundant via studying the gradient saliency, as defined in~\ref{app:B}.

\subsection{Quantum baseline: 1P1Q}
We reproduced the 1P1Q encoding and Variational Quantum Circuit (VQC) Ansatz proposed on the original paper~\cite{bal2025one} to assess the performance against present QML approaches for jet substructure. To match the same setting as the QTTN, we utilize $N=7$ qubits; each of those represents the $p_T$ ordered constituents of the jet, mapped to the Bloch sphere via:
\begin{equation}
\begin{split}
f \cdot \frac{p_T}{p_{T,jet}} \cdot (\eta - \eta_{jet}) \to \theta \\
f \cdot \frac{p_T}{p_{T,jet}} \cdot (\phi - \phi_{jet}) \to \varphi
\end{split}
\end{equation}
where $f$ is a deformation factor parametrized in terms of a learnable variable $w$: $f = 1+2\pi/(1+\exp(-w))$. 
After the RX RY gates encoding of the features, a CNOT ring is followed by $Rot$ gates and measurement on the first qubit, resulting on $3\times N+2$ trainable parameters.
A short discussion about the gradient saliency of this circuit is shown in~\ref{app:B} as well.

\subsection{Classical Baselines}
\label{subsec:baselines}

The performance of the QTTN is benchmarked against three classical machine learning algorithms: a BDT, representing the standard multivariable approach in high-energy physics, a MLP of similar parameter count as the QTTN, and LundNet~\cite{dreyer2021jet}, a successful GNN explicitly designed to process Lund Jet Plane geometries and directly mapping the topological baseline. The input information is the same for each method in all benchmark experiments (except for LundNet-5, as explained).

\subsubsection{Boosted Decision Tree}
\label{subsubsec:bdt}
As a robust benchmark, we employ a standard Gradient Boosted Decision Tree using the \texttt{XGBoost} framework \cite{chen2016xgboost}. For the BDT, the topological structure of the jet cascade is flattened into a one-dimensional array of $N \times 2 = 14$ contiguous classical features. Empty nodes within the sequence are retained as identically zero entries to maintain a consistent feature shape. The hyperparameters were tuned to mitigate overtraining on the relatively sparse input vector, utilizing an ensemble of $N_{\text{est}} = 300$ trees with a maximum depth of 4, amounting to around $13\times 10^3$ parameters and a learning rate of $\eta = 0.05$. 
Two BDTs with identical setting were trained: one on the QTTN inputs and another on the 1P1Q VQC inputs. The relative performance allows assessing the degree of representational expressivity offered by the high dimensional Hilbert space of quantum methods with respect to classical ensemble-based methods, abstracting it from the classical feature extraction of its inputs (the mapping of the Lund Plane and the mapping the particles on the Bloch sphere).

\subsubsection{Multi Layer Perceptron}
In order to assess the QTTN in a complexity versus performance space, we compare the QTTN against a MLP. To ensure a rigorous comparison, the MLP architectures were chosen to have same or higher parameter count as the quantum circuit.

The input data was flattened into a one-dimensional vector $\mathbf{x} \in \mathbb{R}^d$. The MLP utilizes a feed-forward architecture where the number of hidden units was adjusted according to the depth of the quantum circuits, as follows:
\begin{itemize}
    \item For a 1-layer depth ($L=1$), a hidden layer of size (2) was used, totalling 33 parameters.
    \item For a 3-layer depth ($L=3$), a hidden configuration of (4, 3) was used, totalling 79 parameters.
    \item For 5 and 10 layers, hidden sizes of (9) and (17) and parameter count 145 and 273 were employed, respectively.
\end{itemize}
The specific choices of architectures were retained as the most competitive ones among those with similar parameter count. 

\subsubsection{LundNet Graph Neural Network}
\label{subsubsec:lundnet}

To provide a direct topological baseline, we compare the QTTN to a graph-based neural network mapped to the same physical structure. We implement both the full version and a variant of LundNet. 

The LundNet baseline is constructed utilizing the framework provided by original authors \cite{lundnet_code}, which is based on the Deep Graph Library, which in turn uses PyTorch as backend. The full version (hereafter referred to as LundNet-5) processes the full Lund graph using 5 variables per node, namely [$\ln(\Delta R)$, $\ln(z)$, $\ln(k_t)$, $\ln(m)$, $\psi$] (where, given pseudojets $a$ and $b$, and $y_{a,b}$ their rapidity, $m$ is their invariant mass and $\psi = \tan^{-1}(\frac{y_b - y_a}{\phi_b - \phi_a})$).  The modified version of the network (hereafter referred to as LundNet-2) takes as input the same information as QTTN, i.e. the Lund graph (featuring nodes satisfying $\ln(k_t)>1 $) up to the third level of depth, and [$\ln(\Delta R)$, $\ln(z)$] as node features. Due to the dynamic nature of GNNs, missing nodes up to the third level of depth are simply not included in the LundNet-2 input graph, instead of being "zero" encoded as in the QTTN implementation.

LundNet-2 and LundNet-5 are implemented with the same architecture, which corresponds to the one presented in the original paper. The core of the architecture relies on dynamically generating graph representations via Edge Convolution (\texttt{EdgeConv}) blocks \cite{Wang_2019_DGCNN}. 
Both our LundNet implementations comprise six consecutive \texttt{EdgeConv} layers with feature dimensions incrementally expanding up to 128 channels. To prevent vanishing gradients across the deep architecture, shortcut connections are applied across each block. Following the message-passing phase, the node representations are concatenated, passed through a feature fusion layer with Batch Normalization, and collapsed into a single jet-level representation via global mean pooling. The final classification logit is produced by a fully connected layer employing Dropout to regularize the network.  This results in 391430 and 391724 trainable parameters for LundNet-2 and LundNet-5, respectively. 

Comparing the QTTN directly to LundNet implementations allows us to isolate the impact of the quantum Hilbert space and variational entanglement, given an identical physical graph and node features.

\section{Training Strategy}
\label{sec:training}
The optimization of parameterized quantum circuits and deep classical neural networks on sparse physics data requires careful hyperparameter tuning and learning rate management. To ensure a rigorous comparison, all models, the (QTTN), the LundNet GNN, and the BDT, were trained on identically processed subsets of the simulated collision data.
The training was performed on the UniNuvola cloud infrastructure provided by the University of Perugia, utilizing a multi-instance GPU (MIG) slice of an NVIDIA A30 Tensor Core GPU with 24 GB of VRAM. The quantum simulations were executed using the \texttt{PennyLane} framework with the \texttt{JAX} backend, allowing for efficient vectorized computations and automatic differentiation across the entire quantum-classical hybrid architecture.

\subsection{Dataset and Preprocessing}
\label{subsec:dataset_prep}



We summarize the data used for this study and the statistics of the training splits and the generator choice in Table~\ref{tab:dataset}.
\begin{table*}[ht]
\centering
\caption{Overview of the datasets used and the training splits.}
\label{tab:dataset}
\begin{tabular}{lrrrc} 
\toprule
Dataset Name & Train Size & Val Size & Test Size & Generator \\
\midrule
JetGame $W$        & 500,000     & 50,000    & 50,000  & \texttt{Pythia}   \\
JetGame Top        & 500,000     & 50,000    & 50,000   & \texttt{Pythia} \\
JetGame QCD        & 500,000     & 50,000    & 50,000   & \texttt{Pythia} \\
$W^+W^-$ tree level     & 196,000     & 42,000    & 24,000  & \texttt{Pythia}  \\
$W^+W^-$ tree level     & 98,000     & 21,000    & 21,000  & \texttt{Herwig} Dipole  \\
$W^+W^-$ tree level     & 98,000     & 21,000    & 21,000  & \texttt{Herwig} Angular order  \\
$G_{\text{bulk}}\to W^+W^-$     & 196,000     & 42,000    & 42,000  & \texttt{Pythia}  \\


\bottomrule
\end{tabular}
\end{table*}

As described in Section \ref{subsec:lundplane}, zero-safety was applied. 
We note that no manual feature standardization was applied to the Lund Jet Plane coordinates. The raw logarithmic values $x_1$ and $x_2$ were fed directly into the models. For the classical baselines, this provides a raw benchmark. For the QTTN, this design choice tests the efficacy of the learnable stereographic projection: the stretch parameters $\lambda$ and $\omega$ dynamically learn the optimal normalization to map the raw, unscaled QCD phase space onto the normalized surface of the Bloch sphere during the training process.

\subsection{QTTN Optimization}
\label{subsec:qttn_training}

The QTTN was implemented and simulated utilizing the \texttt{PennyLane} quantum machine learning framework \cite{bergholm2018pennylane} paired with the \texttt{JAX} \cite{jax2018github} numerical backend. The integration of \texttt{JAX} allows for the end-to-end Just-In-Time (JIT) compilation of the entire quantum circuit, enabling efficient, vectorized batch processing over the simulated quantum state vectors. 

The quantum model possesses a compact parameter space: the stretch parameters $\lambda_i$ and $\omega_i$, the $L \times N_{\text{edges}}$ CRY parameters $w$, the $(L-1) \times N_{\text{nodes}} \times 3$ local rotation angles $(\alpha, \beta, \gamma)$, the 3 additional rotations on the read-out qubit, and the final linear scaling weights. These parameters were updated by evaluating the binary cross-entropy (BCE) loss on the output logits. 

The optimization was driven by the AdamW algorithm \cite{loshchilov2017decoupled} via the \texttt{Optax} library \cite{deepmind2020jax}, employing a weight decay of $10^{-4}$ to regularize the classical scaling weights. To mitigate the risk of converging to local minima, a cosine annealing learning rate schedule with warmup was utilized. The learning rate was initialized at $10^{-3}$, warmed up to a peak value of $5 \times 10^{-3}$ over the first 10 epochs, and subsequently decayed following a cosine curve down to $10^{-3}$ over the remaining epochs. The QTTN was trained for $E = 50$ epochs utilizing a batch size of $B = 1024$ unless specified otherwise. Identical setting was used for the 1P1Q VQC.

\subsection{Classical Baselines Optimization}
\label{subsec:classical_training}

The classical baselines BDT and MLP were optimized to ensure they reached their maximum theoretical performance on the $N=7$ node graphs. As for LundNet, both LundNet-2 and LundNet-5 training procedures were implemented with the very same methods and hyperparameters as in the original LundNet-5 implementation.

\textbf{LundNet:} the GNN training was implemented using the framework provided by original authors \cite{lundnet_code}, with a batch size of $B = 256$ for $E = 30$ epochs. The CrossEntropy loss was minimized using the Adam optimizer \cite{Kingma:2014aa}, with a learning rate initialized at $10^{-3}$ and decreased by a factor of 0.1 every 10 epochs. To prevent overtraining, Dropout layers with a rate of $0.1$ were actively utilized during the training phase. The model achieving the best validation accuracy is retained as the best model and tested on the test set.

\textbf{BDT:} the Boosted Decision Tree was trained utilizing the \texttt{XGBClassifier} API. The depth of the trees was restricted to 4 to prevent the algorithm from memorizing the sparse arrays. The ensemble was grown to 300 estimators with a learning rate of $0.05$. The BDT was optimized directly on the flattened training samples.

\textbf{MLP:} the Multi Layer Perceptron employs the hyperbolic tangent ($\tanh$) activation function and is optimized using the Adam stochastic optimizer  with an initial learning rate of $10^{-3}$. The network was trained for 50 epochs using mini-batch gradient descent with a patience-based early stopping criterion of 20 iterations to prevent overfitting. Convergence was ensured monitoring the Binary CrossEntropy loss and validation AUC.

\section{Results and Discussion}
\label{sec:results}


We evaluate models based on critical properties for the HEP landscape: 
\begin{itemize}
    \item Performance on the entire datasets in~\ref{subsec:performance_all},
    \item Performance with partial dataset for low-data regime~\ref{subsec:performance_lowdata},
    \item Out-of-Distribution generalization performance on alternative hadronization and showering models~\ref{subsec:performance_herwigs},
\end{itemize}

Finally, we produce a dedicated performance study on polarization tagging for the VBS topology to fully exploit the different production dynamics in~\ref{subsec:vbs}.

We quantify the discriminative power of the models using the Area Under the Receiver Operating Characteristic Curve (AUC). Its statistical uncertainty is evaluated via the DeLong method~\cite{delong1988comparing}; the values are found to be comparable with bootstrap method and confirmed further with Cross Validation $k-$fold.

\subsection{Performance}\label{subsec:performance_all}
We show the classification performance in terms of the AUC for all the dataset, quantum and classical ML methods in Table~\ref{tab:model_summary}.  The QTTN model is additionally shown with four different choices for the number of layers $L$: 1, 3, 5 and 10, corresponding to 25, 79, 133 and 268 independent trainable parameters respectively.
For W tagging, 10 layers are needed to outperform the BDT (13 thousand parameters) trained on the same input, and 5 layers to match the MLP with similar parameter count. While the LundNet-2 clearly outperforms QTTN, we observe that this method is vastly superior to the other quantum method 1P1Q.

For top tagging, we reproduce the excellent performance already shown by 1P1Q in~\cite{bal2025one}. However, we observe that the classical BDT trained on the same 1P1Q input features significantly outperforms the 1P1Q quantum circuit, suggesting that the model's discriminative power originates primarily from the information content of the classical encoding rather than the expressive capacity of the VQC.
The QTTN is found to match the performance of the classical BDT on its input with 10 layers, lacking behind LundNet-2.

In both polarization tagging scenarios we observe a stronger match of the QTTN to the classical baselines, and outperform 1P1Q by large margin. In $G_\text{bulk} \to W^+W^-$, we observe that the QTTN with 5 layers was able to match the LundNet-2.

\begin{figure}[htpb]
    \centering
    \includegraphics[width=0.45\textwidth]{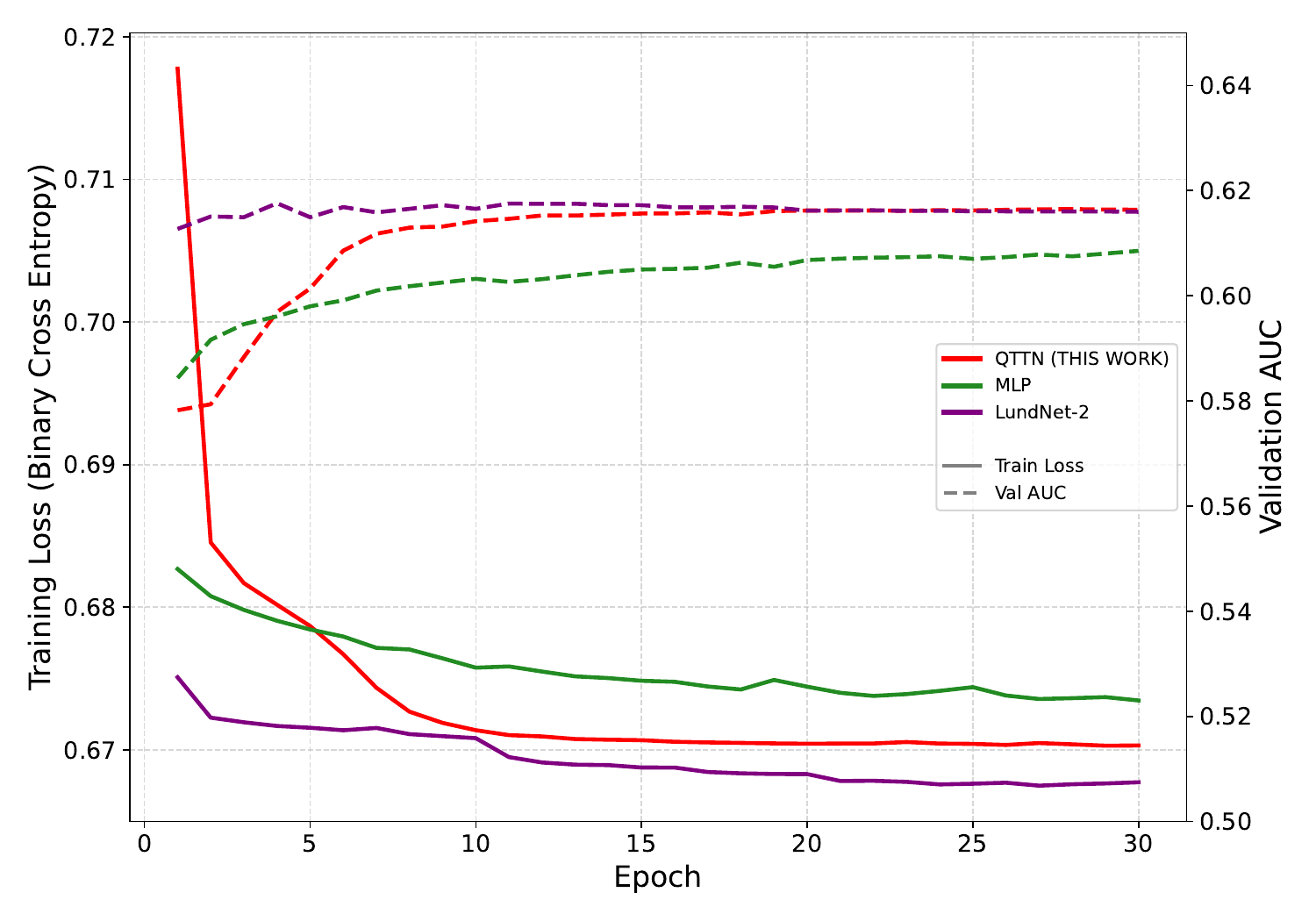}
    \caption{ Training dynamics of the QTTN $L=3$ for the $G_\text{bulk}\to W^+W^-$ process and selected benchmarks over the first 30 epochs, showing the stable minimization of the Binary Cross-Entropy loss alongside the corresponding validation AUC. }
    \label{fig:training_dynamics}
\end{figure}

The training dynamics is shown in Figure~\ref{fig:training_dynamics} for a representative scenario. 

\begin{table*}[htpb]
\centering
\caption{Summary of model configurations and their corresponding AUC performance evaluated across different tagging datasets.}
\label{tab:model_summary}
\begin{tabular}{l|l|c|c}
    \hline
    \textbf{Model} & \textbf{Configuration} & \textbf{AUC} & \textbf{Dataset} \\
    \hline
    QTTN (THIS WORK)                 & 1 Layer, 25 params         & $0.706 \pm 0.002$    & W tagging       \\
    MLP                   & 33 params                 & $0.728 \pm 0.002$    &         \\
    QTTN                  & 3 Layers, 79 params         & $0.795 \pm 0.001$    &         \\
    MLP                   & 79 params                 & $0.806 \pm 0.001$    &         \\
    QTTN                  & 5 Layers, 133 params        & $0.813 \pm 0.001$    &         \\
    MLP                   & 145 params                & $0.814 \pm 0.001$    &        \\
    QTTN                  & 10 Layers, 268 params       & $0.827 \pm 0.001$    &        \\
    MLP                   & 273 params                & $0.826 \pm 0.001$    &        \\
     \hline
    BDT (QTTN inputs)                   & 13k params                   & $0.821 \pm 0.001$    &         \\
    LundNet-2              & 391k params        & $\mathbf{0.844} \pm 0.001$    &        \\
    1P1Q                  & 23 params                  & $0.663 \pm 0.002$    &        \\
    BDT (1P1Q inputs)     & 13k params                   & $0.754 \pm 0.002$    &       \\
    \hline \hline
    QTTN (THIS WORK)                 & 1 Layer, 25 params         & $0.804 \pm 0.001$    & Top tagging     \\
    MLP                   & 33 params                 & $0.858 \pm 0.001$    &      \\
    QTTN                  & 3 Layers, 79 params         & $0.886 \pm 0.001$    &       \\
    MLP                   & 79 params                 & $0.883 \pm 0.001$    &       \\
    QTTN                  & 5 Layers, 133 params        & $0.893 \pm 0.001$    &       \\
    MLP                   & 145 params                & $0.891 \pm 0.001$    &       \\
    QTTN                  & 10 Layers, 268 params       & $0.898 \pm 0.001$    &       \\
    MLP                   & 273 params                & $0.897 \pm 0.001$    &       \\
     \hline
    BDT (QTTN inputs)                  & 13k params                   & $0.896 \pm 0.001$    &      \\
    LundNet-2              & 391k params         & $0.905 \pm 0.001$    &       \\
    1P1Q                  & 23 params                  & ${0.913} \pm 0.001$    &       \\
    BDT (1P1Q inputs)     & 13k params                   & $\mathbf{0.935} \pm 0.001$    &       \\
    \hline \hline
    QTTN (THIS WORK)                  & 1 Layer 25 params         & $0.616 \pm 0.003$    &  Polar. $W^+W^-$     \\
    MLP                   & 33 params                 & $0.614 \pm 0.003$    &       \\
    QTTN                  & 3 Layers, 79 params         & $0.639 \pm 0.003$    &       \\
    MLP                   & 79 params                 & $0.629 \pm 0.003$    &       \\
    QTTN                  & 5 Layers, 133 params        & $0.639 \pm 0.003$    &       \\
    MLP                   & 145 params                & $0.635 \pm 0.003$    &       \\
    QTTN                  & 10 Layers, 268 params       & $0.640 \pm 0.003$    &       \\
    MLP                   & 273 params                & $0.637 \pm 0.003$    &       \\
     \hline
    BDT (QTTN inputs)                 & 13k params                   & $0.639 \pm 0.003$    &       \\
    LundNet-2              & 391k params         & $\textbf{0.642} \pm 0.003$    &       \\
    1P1Q                  & 23 params                  & $0.553 \pm 0.003$    &      \\
    BDT (1P1Q inputs)     & 13k params                    & $0.621 \pm 0.003$    &      \\
    \hline \hline
    QTTN (THIS WORK)                  & 1 Layer, 25 params         & $0.592 \pm 0.003$    &   Polar. $G_{\text{bulk}}\to W^+W^-$    \\
    MLP                   & 33 params                 & $0.598 \pm 0.003$    &       \\
    QTTN                  & 3 Layers, 79 params         & $0.615 \pm 0.003$    &       \\
    MLP                   & 79 params                 & $0.608 \pm 0.003$    &       \\
    QTTN                  & 5 Layers, 133 params        & $\textbf{0.616} \pm 0.003$    &       \\
    MLP                   & 145 params                & $0.612 \pm 0.003$    &       \\
    QTTN                  & 10 Layers, 268 params       & $\textbf{0.616} \pm 0.003$    &       \\
    MLP                   & 273 params                & $0.613 \pm 0.003$    &       \\
     \hline
    BDT (QTTN inputs)                  & 13k params                    & $\textbf{0.616} \pm 0.003$    &       \\
    LundNet-2      & 391k params         & $\textbf{0.616} \pm 0.003$    &       \\
    1P1Q                  & 23 params                  & $0.569 \pm 0.003$    &      \\
    BDT (1P1Q inputs)     & 13k params                    & $0.608 \pm 0.003$    &      \\
    \hline
\end{tabular}
\end{table*}

\begin{figure*}[htbp]
     \centering
     \begin{subfigure}[b]{0.45\textwidth}
         \centering
         \includegraphics[width=\textwidth]{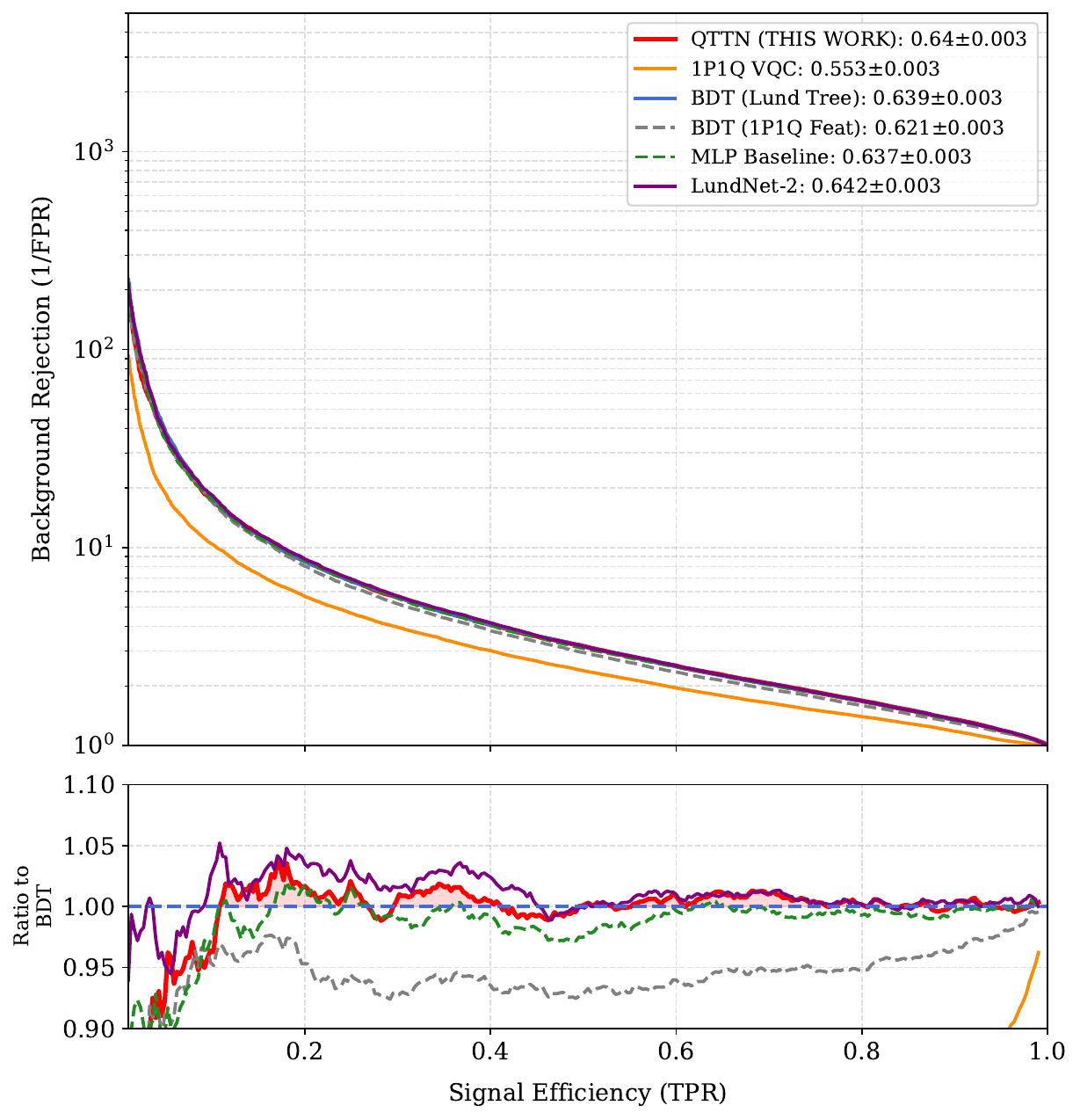}
         \caption{Performance on $W^+W^-$ polarization}
     \end{subfigure}
     \hfill
     \begin{subfigure}[b]{0.45\textwidth}
         \centering
         \includegraphics[width=\textwidth]{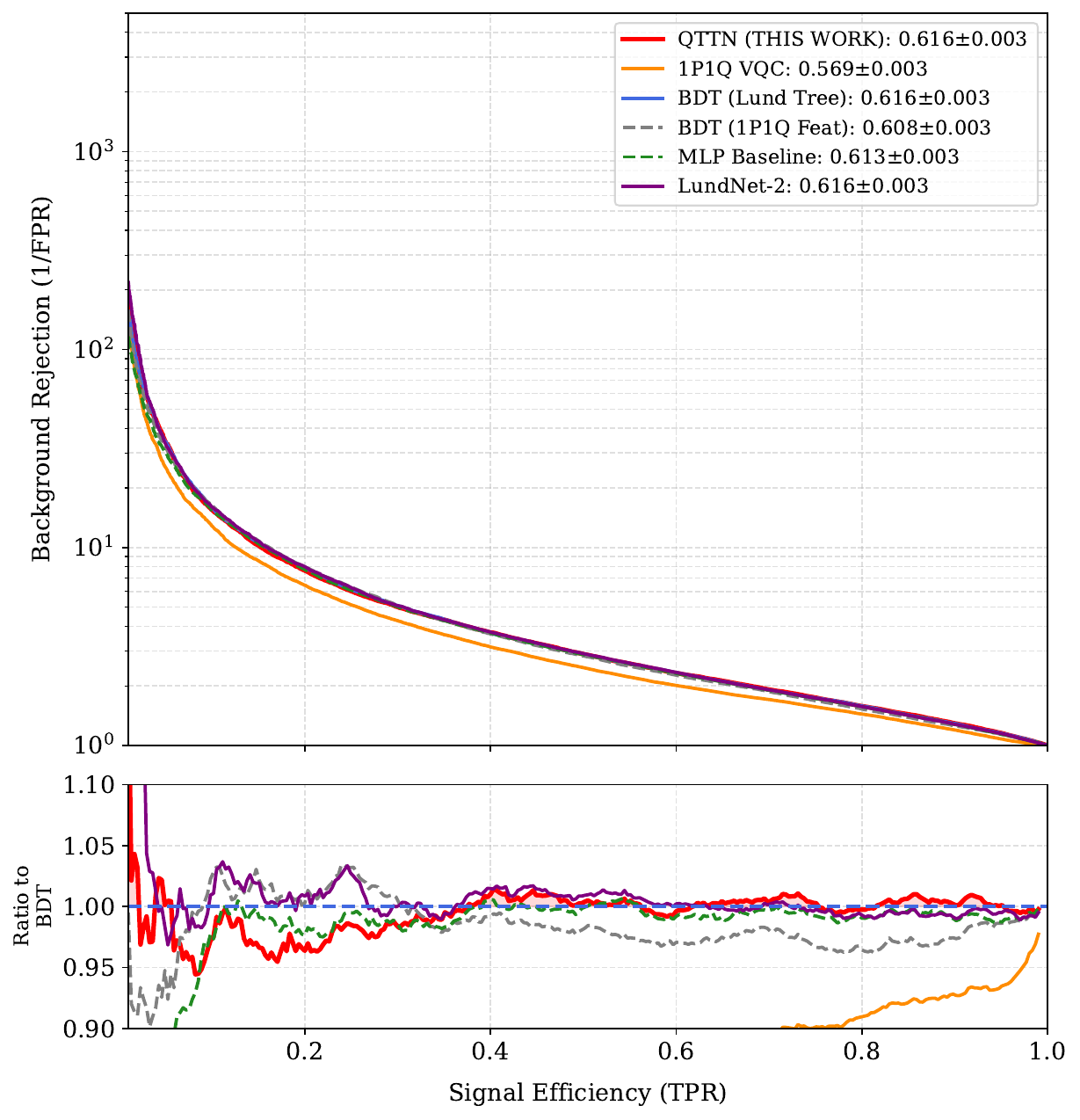}

         \caption{Performance on $G_{\text{bulk}}\to W^+W^-$ polarization}
     \end{subfigure}

     \vspace{10pt} 

     \begin{subfigure}[b]{0.45\textwidth}
         \centering
         \includegraphics[width=\textwidth]{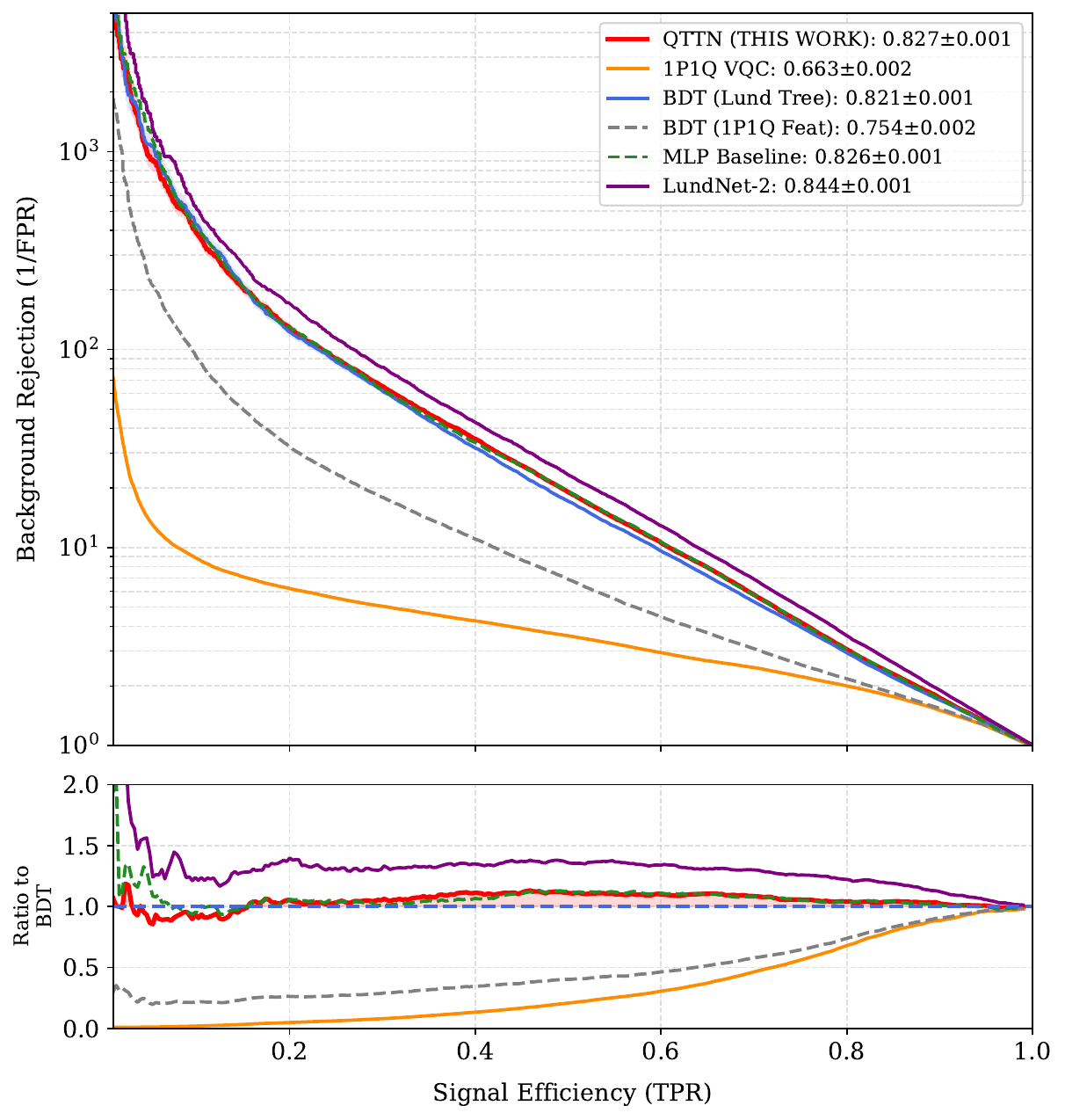}

         \caption{Performance on $W$ tagging}
     \end{subfigure}
     \hfill
     \begin{subfigure}[b]{0.45\textwidth}
         \centering
         \includegraphics[width=\textwidth]{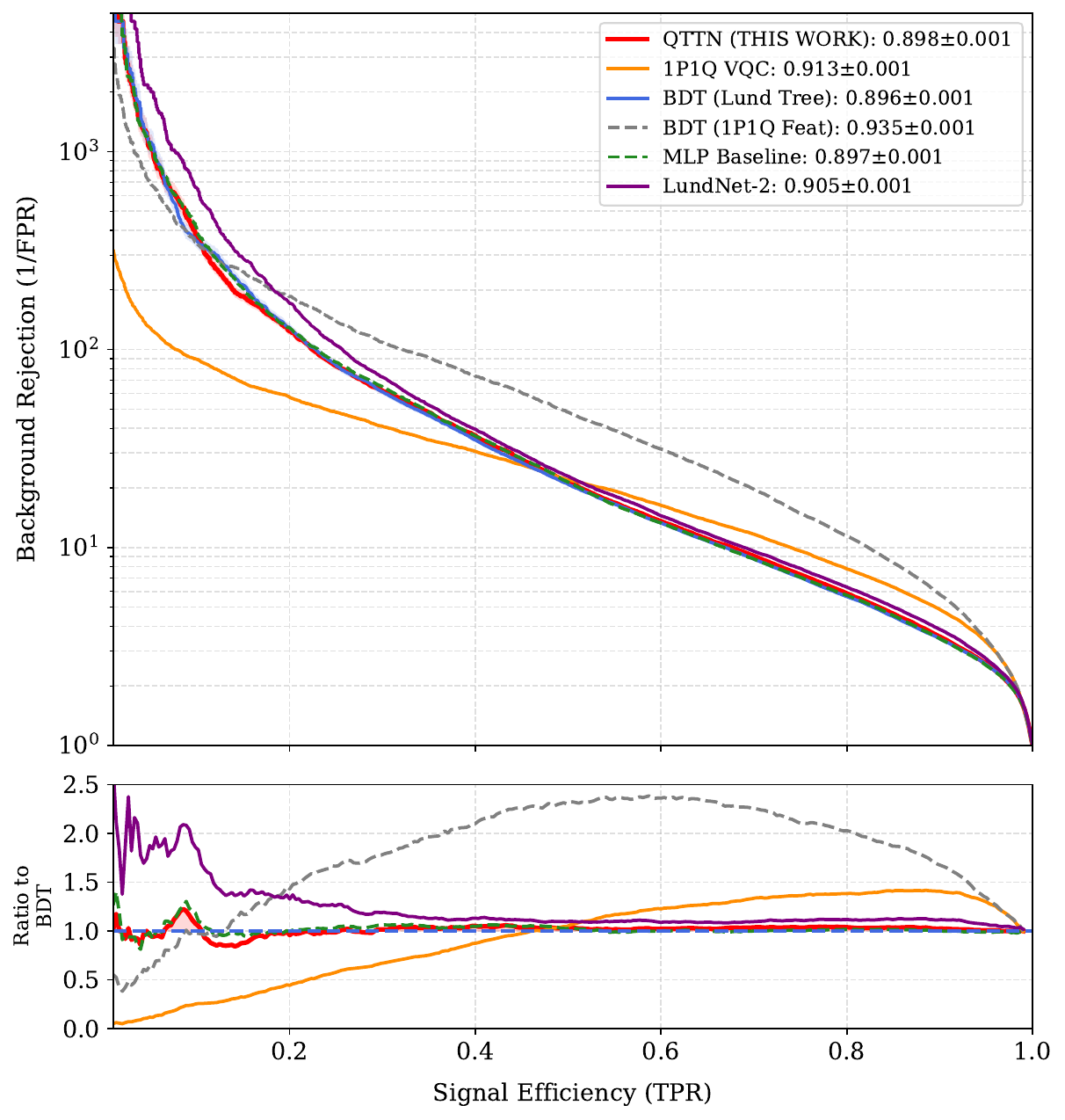}

         \caption{Performance on Top Tagging}
     \end{subfigure}
     
     \caption{Comparison of ROC curves for the QTTN and the classical and quantum benchmarks considered in this work, for the four scenarios considered.}
     \label{fig:QTTN10_ROC}
\end{figure*}

The ROC curves are shown in Figure~\ref{fig:QTTN10_ROC} for the $L=10$ QTTN configuration; for the different $L$ choice the ROC curves are shown in the Appendix, Figure~\ref{fig:QTTN_ROC_LAYERS}.

\begin{figure*}[htpb]
    \centering
    \includegraphics[width=0.95\textwidth]{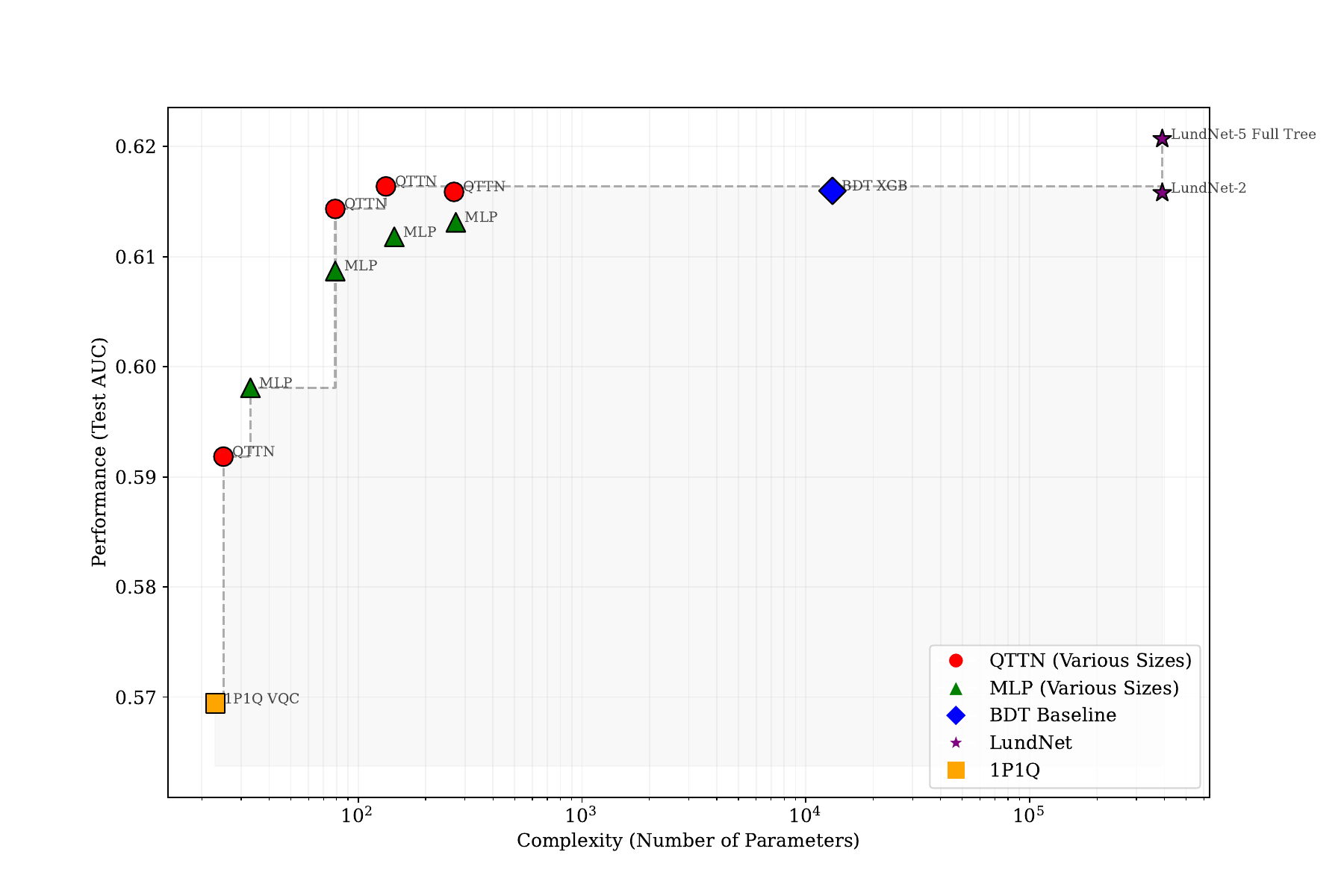}
    \caption{The Pareto front showing complexity (log scale) versus performance, showing the QTTN (in red) defined in this work with four different layer choices ($L=1,3,5$ and $10$), against MLP baseline of similar parameter count as the QTTN, the BDT trained on QTTN inputs and the LundNet-2 and LundNet-5. QTTN pushes the Pareto front in the low parameter count region. The dataset used is the polarized graviton scenario. }
    \label{fig:pareto}
\end{figure*}

The Pareto front plot comparing performance and complexity of each model is shown in Figure~\ref{fig:pareto}.

\subsection{Low-data regime}\label{subsec:performance_lowdata}
In many analyses, background and signal processes are often modeled via Monte Carlo (MC) simulations, allowing (Q)ML models to potentially train on very large datasets. However, high-fidelity, full simulations of the detector require vast amounts of computing time, sometimes leading to sizable statistical uncertainty in measurements~\cite{Albrecht:2019aa}.  
Additionally, e.g. high-order QCD processes are difficult to model accurately with MC event generators (e.g.~\cite{Chisholm:2021aa,Collaboration:2017aa}), making data-driven methods the preferred choice for estimating yields in complex final states. In searches for highly boosted particles or rare BSM signatures, the kinematic cuts are so extreme that signal regions are typically populated with few events. 
As a consequence, many analyses rely on statistical control regions or limited MC samples to train classifiers, forcing ML models to operate and remain stable in regimes where the available training data are intrinsically sparse: the low-data regime. This makes the development of architectures, regularization strategies and uncertainty-aware training procedures that remain performant in the low-data regime an important direction~\cite{Karagiorgi:2021aa,Kasieczka:2021aa,Hallin:2025aa,Beauchesne:2023aa}.

In this context, QML has been shown in recent years to be a promising alternative to potentially reduce the amount of training data required, thanks to the expressivity provided by the exponentially large Hilbert space, achieving faster generalization and maintaining stability even when trained on a fraction of the data typically required for deep learning models~\cite{Caro:2021aa,Sakhnenko:2025aa,Reese:2022aa}.

To demonstrate this in jet substructure, we train the QTTN on the polarized $W^+W^-$ sample and $W$ tagging, reducing the number of training events by one order of magnitude for each point, while maintaining the entire statistics for the test set. To ensure statistical accuracy, we use a k-fold approach, k=10, sampling the training and validation set from the available data.

\begin{figure*}[t] 
    \centering
    \begin{subfigure}[b]{0.48\textwidth}
        \centering
        \includegraphics[width=\textwidth]{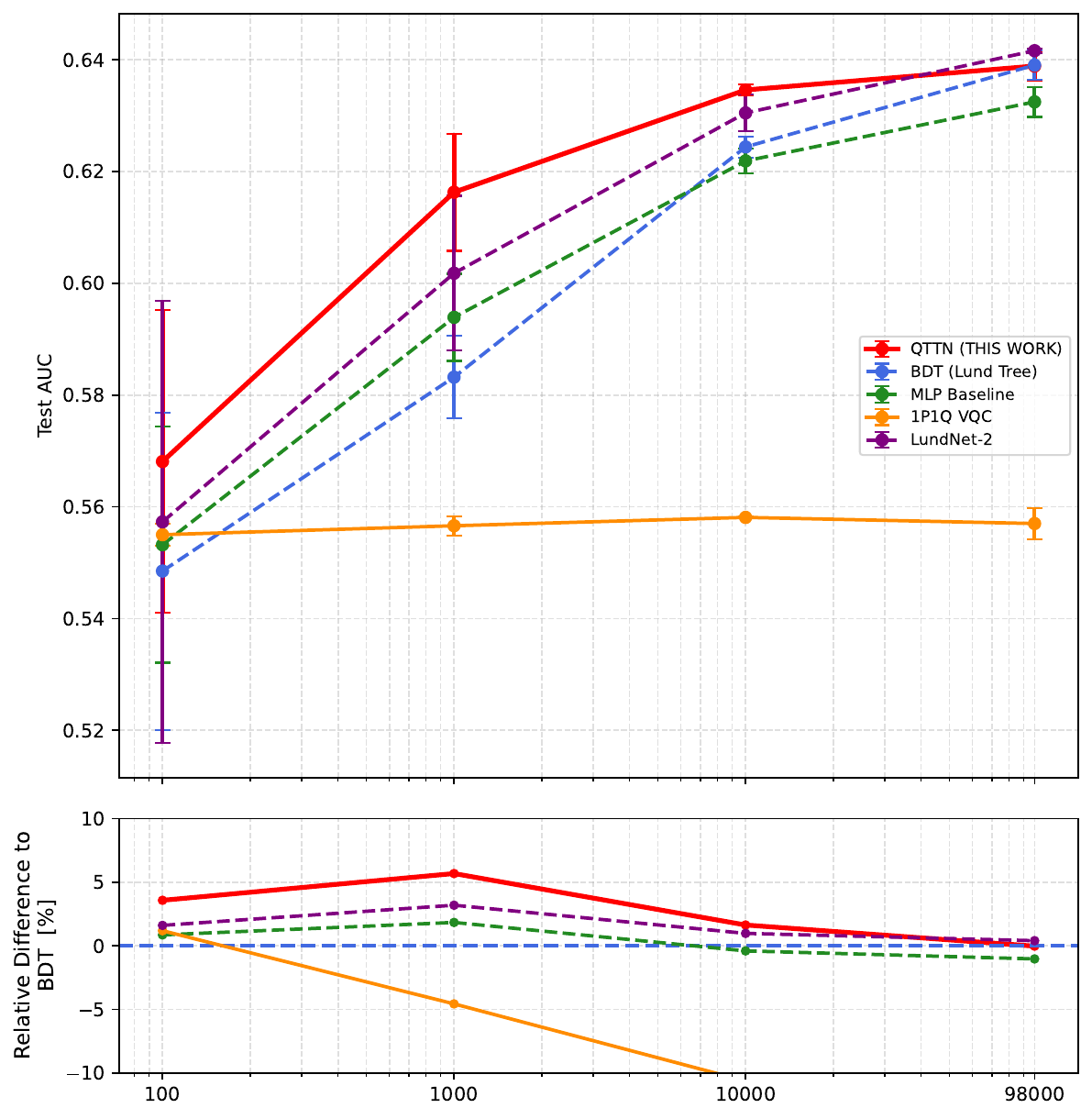}
        \caption{AUC score on the test set as a function of the training data statistics per class, for the $W^+W^-$ tree level sample for the models considered. The red line corresponds to the QTTN with $L=3$ proposed in this work.}
        \label{fig:lowdatawwborn}
    \end{subfigure}
    \hfill
    \begin{subfigure}[b]{0.48\textwidth}
        \centering
        \includegraphics[width=\textwidth]{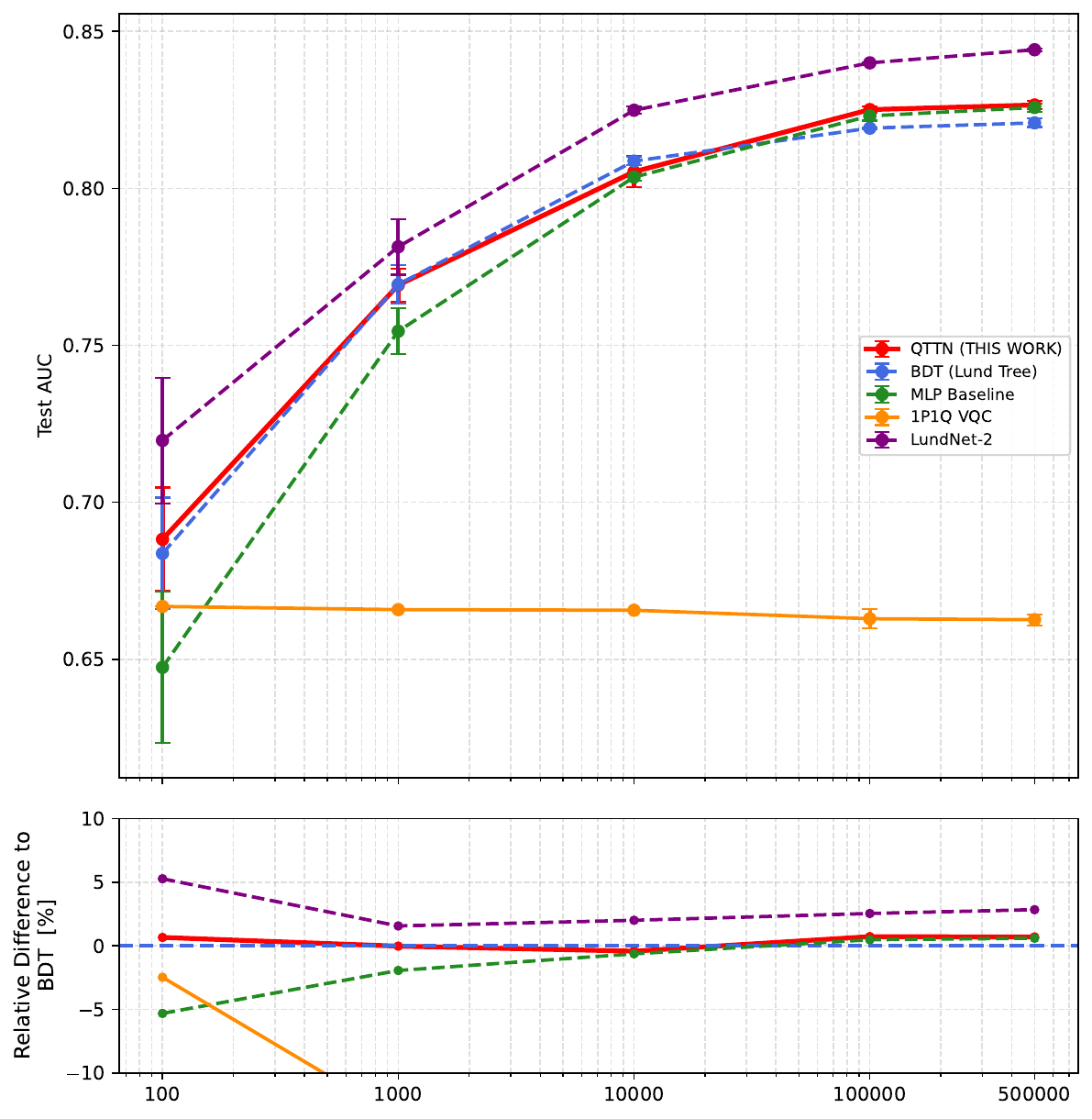}
        \caption{AUC score on the test set as a function of the training data statistics per class, for the $W$ vs QCD sample for the models considered. The red line corresponds to the QTTN with $L=10$ proposed in this work.}
        \label{fig:lowdatawtagging}
    \end{subfigure}
    \caption{Comparison of AUC scores and relative differences for different training data statistics across the considered models. The bottom panel shows the percentage relative difference with respect to the BDT trained on the same input as the QTTN.}
    \label{fig:combined_results}
\end{figure*}

The result is summarized in Figure~\ref{fig:lowdatawwborn} as a function of the Test AUC for the polarized sample ($WW$ tree level), and in Figure~\ref{fig:lowdatawtagging} for the W vs QCD classification. In the former, the classical methods, including the large model LundNet-2, show a more pronounced dependence on the size of the training dataset, while QTTN demonstrates a smoother performance degradation dynamics. In particular, the QTTN trained on 1000-10000 events shows similar performance as for models trained with an order of magnitude more data, indicating significant gains in this regime, while the variance of the test does not allow conclusive statement in the 100 events region.

In the W tagging task, where the QTTN struggles more against LundNet-2, we can draw some interesting conclusions: (i) LundNet-2 shows a faster performance degradation against QTTN, albeit keeping its competitive advantage; (ii) the BDT matches or beats the performance of the QTTN in the 100-10000 data region, but underperforms in the high statistical region (100000 events onwards); (iii) the MLP of similar parameter count as the QTTN ties in the high data region, but shows a significant degradation in performance towards the low data region, underperforming against the QTTN.

Even if both QTTN and LundNet-2 are models based on the strong physics bias of the Lund tree declustering history, the small parameter count of the quantum architecture acts as natural regularizer, preventing overfitting that typically plagues high capacity classical networks in the presence of limited statistics. This suggests that the quantum circuit expressive power is better aligned with the underlying symmetry of the classification task. These results highlight the potential of quantum and quantum-inspired architectures as a robust alternative for HEP applications where high quality labeled data may be computationally expensive or physically limited.

\subsection{Domain Transfer Generalization: \texttt{Pythia} vs \texttt{Herwig}}\label{subsec:performance_herwigs}
We investigate the Out-of-Distribution (OOD) generalization capability of the QTTN model compared with the classical baseline; in particular, we want to judge the capability of the model not to implicitly overfit the specific hadronization and parton shower model while retaining discrimination capability. This is of great interest in jet substructure, as radically different performances of a (quantum) machine learning method trained on one generator and tested on another are directly connected to a high systematic uncertainty.
In e.g. searches for new physics, analyses are typically challenged by vast amounts of QCD background. In this context, a tagger with superior AUC but higher associated systematic uncertainties could be outperformed in signal significance $Z$ by a method with inferior AUC but smaller associated systematic uncertainties. From the asymptotic expression~\cite{cowan2011asymptotic} we see that the impact of this uncertainty scales quadratically both as a function of background yield and as a function of the relative uncertainty:
$$Z = \frac{s} {\sqrt{b+\sigma^2_b}} + \mathcal{O}(s/b) + \mathcal{O}(\sigma^2_b/b)$$
where $s$ is the signal yield, $b$ the background yield and $\sigma_b$ the absolute background systematic uncertainty, or $\sigma_b=b\times\sigma_{rel}$, with $\sigma_{rel}$ the relative uncertainty in percentage. 
It has been shown~\cite{sahu2024ml} that machine learning taggers which make use of low level input, such as tracks or calorimeter towers, exhibit a hadronization and parton showering systematic uncertainty reaching as high as 40\% for CNN/GNN classifier on top tagging, while classifiers based on High-Level Features such as jet mass and N-subjettiness remain largely unaffected by the choice of the generator.
In this sense the trade-off between performance and uncertainty should be carefully evaluated based on the specific analysis and background levels in the signal region.

The Lund Jet Plane does not make use of individual constituent susceptible of a high degree of mismodeling; however we argue that large networks such as LundNet could still inadvertently exploit soft and non-perturbative regions which are generator-dependent.


To isolate the true degree of model overfitting, we measure the relative transfer gap between models. For a target domain $T \in \{\texttt{Pythia}, \texttt{Herwig}\}$, we compare the performance of a natively trained model ($\text{AUC}_{T \to T}$) against the transferred performance of a model trained on the alternate generator ($\text{AUC}_{S \to T}$). The gap is defined as:
\begin{equation}
\Delta_{S \to T} = \frac{\left| \text{AUC}_{T \to T} - \text{AUC}_{S \to T} \right|}{\text{AUC}_{T \to T}}
\end{equation}

The cross-evaluations are summarized in Table~\ref{tab:transfer_gap_full}. Crucially, the results demonstrate that the raw $\sim 4-5\%$ drop in AUC observed when moving from \texttt{Pythia} to \texttt{Herwig} is almost entirely attributable to the inherent features of the \texttt{Herwig} shower model, not domain overfitting. For example, the QTTN trained on \texttt{Pythia} achieves an AUC of $0.609$ on \texttt{Herwig}. A QTTN trained natively on \texttt{Herwig} achieves an almost identical AUC. 

We observe that the QTTN achieves domain transfer with an average transfer gap of just $0.2\%$, similar to other low-parameter methods considered, and much lower than the highly-parameterized LundNet-5 ($\Delta = 0.8\%$) while matching or closely tailing its performance on the hadronization and showering models considered, and outperforming other classical and quantum methods. This supports the hypothesis that while heavier architectures like LundNet extract marginal fractional gains by exploiting non-perturbative, generator-specific phase spaces, the QTTN isolates the universal, underlying physics.

\begin{table*}[htpb]
\centering
\caption{Domain transfer gap across \texttt{Pythia} (P) and \texttt{Herwig} (H) generators, evaluated for both Angular Order and Dipole shower variations. The Native performance (e.g., trained and tested on Pythia) acts as the theoretical maximum for a given target domain. The transfer gap ($\Delta$) measures the relative degradation when evaluating a model trained on the alternate domain. QTTN is used in the $L=5$ configuration.}
\label{tab:transfer_gap_full}
\renewcommand{\arraystretch}{1.2}
\begin{tabular}{l | c c c | c c c | c}
    \hline\hline
    \multirow{2}{*}{\textbf{Model}} & \multicolumn{3}{c|}{\textbf{Target Domain: Pythia}} & \multicolumn{3}{c|}{\textbf{Target Domain: Herwig}} & \multirow{2}{*}{\textbf{Mean $\Delta$}} \\
    \cline{2-7}
    & Native (P$\to$P) & Trans. (H$\to$P) & $\Delta_{H \to P}$ & Native (H$\to$H) & Trans. (P$\to$H) & $\Delta_{P \to H}$ & \\
    \hline
    \multicolumn{8}{c}{\textbf{Herwig Parton Shower: Angular Order}} \\
    \hline
    QTTN (THIS WORK)      & 0.640 & 0.637 & 0.4\%  & 0.609 & 0.609 & \textbf{0.1\%}  & \textbf{0.2\%} \\
    BDT                   & 0.639 & 0.636 & 0.5\%  & 0.609 & 0.609 & \textbf{0.1\%}  & {0.3\%} \\
    MLP                   & 0.634 & 0.634 & \textbf{0.0\%}  & 0.609 & 0.606 & 0.4\%  & \textbf{0.2\%} \\
    1P1Q                  & 0.549 & 0.556 & 1.4\%  & 0.551 & 0.541 & 1.7\%  & {1.5\%} \\
    LundNet-2       & 0.642 & \textbf{0.639} & 0.5\%  & 0.609 & 0.610 & \textbf{0.1\%}  & {0.3\%} \\
    LundNet-5              & \textbf{0.643} & 0.636 & 1.2\%  & 0.606 & \textbf{0.611} & 0.8\%  & {1.0\%} \\
    \hline
    \multicolumn{8}{c}{\textbf{Herwig Parton Shower: Dipole}} \\
    \hline
    QTTN (THIS WORK)      & 0.640 & 0.638 & 0.3\%  & 0.612 & \textbf{0.614} & 0.2\%  & \textbf{0.2\%} \\
    BDT                   & 0.639 & 0.636 & 0.5\%  & 0.612 & 0.613 & \textbf{0.1\%}  & {0.3\%} \\
    MLP                   & 0.634 & 0.633 & \textbf{0.2\%}  & 0.610 & 0.611 & \textbf{0.1\%}  & \textbf{0.2\%} \\
    1P1Q                  & 0.549 & 0.557 & 1.6\%  & 0.554 & 0.545 & 1.8\%  & {1.7\%} \\
    LundNet-2       & 0.642 & 0.638 & 0.6\%  & \textbf{0.613} & 0.613 & \textbf{0.1\%}  & {0.3\%} \\
    LundNet-5              & \textbf{0.643} & \textbf{0.639} & 0.7\%  & 0.610 & \textbf{0.614} & 0.7\%  & {0.7\%} \\
    \hline\hline
\end{tabular}
\end{table*}

\subsection{VBS Polarization tagging}\label{subsec:vbs}
The LundNet and QTTN polarization models trained with graviton and $pp\rightarrow W^+W^-$ SM samples are then applied to the VBS samples, without flattening the $p_T$ spectrum of the boosted hadronically decaying $W$ jets. 
Because of the presence of two additional jets in the final state, a selection on the event topology is performed, requiring a large invariant mass ($m_{j_1j_2} > 500 $ GeV) and a large rapidity separation of the two anti-$k_t$ $R$=0.4 tagging jets ($|\Delta \eta_{j_1j_2}| > 2.5 $), being both a typical signature of VBS events. At the same time, we require, after that this preselection is applied, exactly two anti-$k_t$ R=0.8 jets in the events. If more than two anti-$k_t$ R=0.4 are found, the pair with the highest dijet mass are considered as the two tagging jets. Results are summarized in Table \ref{tab:PolOutput}, where different models are applied to VBS samples. In all cases the AUC outperforms the previous results obtained from polarization studies in hadronically decaying W boson, without any tuning of the clustering or subjet extraction algorithms \cite{Dey:2021sug}.
\begin{table*}[htpb]
\centering
\caption{Performance of the LundNet and QTTN $L=3$ models tested on jets from purely longitudinal and transverse fully hadronic VBS samples.}
\label{tab:PolOutput}
\begin{tabular}{c|c|l}
    \hline
    \textbf{Model} &  \textbf{Train Dataset} & \textbf{AUC VBS} \\
    \hline
    LundNet-2 & $pp\rightarrow W^+W^-$ & 0.648 $\pm$ 0.001\\
    LundNet-5 & $pp\rightarrow W^+W^-$ &  \textbf{0.664} $\pm$ 0.001\\
    QTTN & $pp\rightarrow W^+W^-$ & 0.644 $\pm$ 0.002 \\
    \hline
    LundNet-2 & RS graviton & 0.656 $\pm$ 0.001\\
    LundNet-5 & RS graviton &  0.655 $\pm$ 0.001\\
    QTTN & RS graviton & \textbf{0.660} $\pm$ 0.002 \\
    \hline
\end{tabular}

\end{table*}
\subsection{The Role of the Stereographic Stretch Parameter}
\label{subsec:lambda_discussion}

The improved performance of the  QTTN architecture is primarily due to the behavior of the differentiable stereographic encoding. As noted in Section~\ref{sec:training}, the raw Lund coordinates were fed into the models without any prior statistical standardization. 

While the classical BDT is naturally invariant to monotonic feature scaling, neural networks (both classical and quantum) are notoriously sensitive to unscaled data. For the QTTN, the learnable stretch parameters $\boldsymbol\lambda$ and $\boldsymbol\omega$ successfully absorbed this preprocessing step; by adjusting during optimization, they identify the optimal scale for projecting raw QCD phase space features directly onto the Bloch sphere.

Furthermore, the ``zero-safe'' property of the mapping behaved exactly as theoretically intended. Empty tree nodes mapped precisely to the $|0\rangle$ state, ensuring that the quantum rotation gates did not inject spurious noise into the latent representation. Preserving this structural sparsity, the QTTN maintained the IRC safety demanded by theoretical QCD.

\subsection{Hardware Feasibility and NISQ Suitability}
\label{subsec:nisq}

Besides its classification performance, the QTTN architecture is structurally well-suited for near-term hardware deployment. Traditional constituent based quantum models, such as the 1P1Q embeddings, require qubit counts that scale linearly with the variable particle multiplicity of the jet, often exceeding 30 to 40 qubits per event. In addition to introducing IRC unsafety from collinear splittings, this approach results in gate depths that are highly susceptible to decoherence on current NISQ hardware.

Abstracting the jet into a fixed-depth Lund Jet Plane tree, our QTTN strictly bounds the qubit requirement to $N=7$. The CRY entanglement strategy follows the C/A declustering topology, requiring only 6 two-qubit CRY per layer rather than an all-to-all entanglement scheme ($O(N^2)$ gates). With only 5 variational layers, the circuit depth remains shallow. The stable convergence observed in Figure~\ref{fig:training_dynamics} highlights that this constrained parameter space mitigates the barren plateau phenomenon, allowing for training and deployment on NISQ devices.

\subsection{QTTN $L=1$ on SpinQ Triangulum quantum hardware}

\begin{figure}[htpb]
    \centering
    \includegraphics[width=0.45\textwidth]{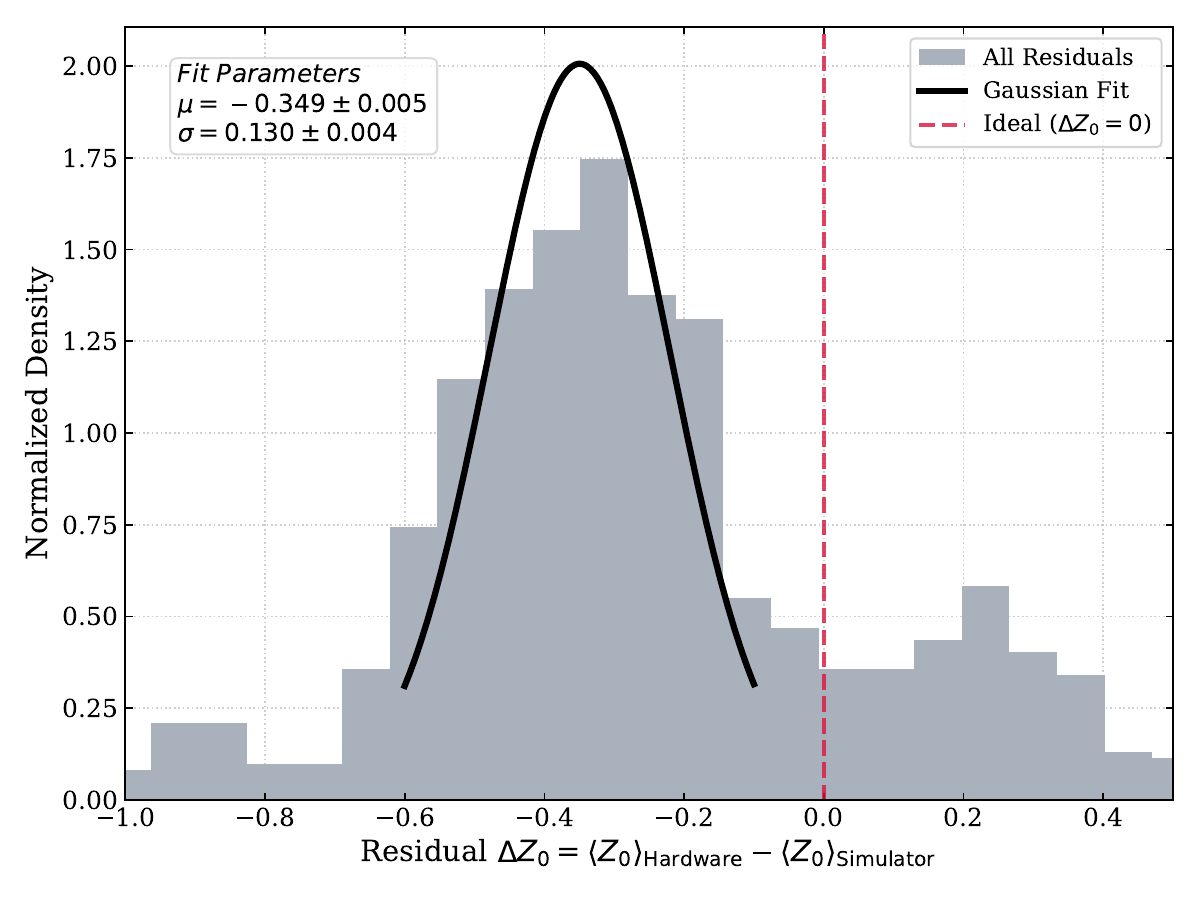}
    \caption{Distribution of residuals between hardware-evaluated and ideal simulator expectation values for the simplified QTTN architecture executed on the SpinQ Triangulum quantum processor for the $pp\to W^+W^-$ polarization tagging. The Gaussian fit yields a mean bias of $\mu = -0.349 \pm 0.005$ and a standard deviation of $\sigma = 0.130 \pm 0.004$. Even though the hardware run shows a systematic shift likely due to relaxation effects, the model retains discrimination power between classes.}
    \label{fig:spinq_results}
\end{figure}

To demonstrate the practical feasibility of the proposed QTTN architecture on current quantum hardware, we implemented a $L=1$ version of the model on the SpinQ Triangulum quantum processor, a 3 qubit liquid-state NMR device using $^{19}F$ nuclear spins as qubits~\cite{feng2022spinq}, designed mainly as an educational platform.
Due to hardware constraints in number of qubits and available gates, we reduced the tree depth to $D=2$, resulting in a binary tree with $N=3$ nodes (root and two children). 
Despite the shallowness, the architecture still captures the essential hierarchical structure of the jet cascade and allows us to validate the core principles of our approach on real quantum hardware. The training was performed using a small subset of around 1000 jets of the polarized $pp\to W^+W^-$ dataset, and the results showed consistent trends with the full QTTN model, confirming the viability of our design for near-term quantum devices. 
In Figure \ref{fig:spinq_results}, we present the distribution of residuals between the hardware-evaluated expectation values and the ideal simulator predictions ($\Delta \langle Z_0 \rangle = \langle Z_0 \rangle_{\text{HW}} - \langle Z_0 \rangle_{\text{Sim}}$). We performed a Gaussian fit to quantifying the systematic and stochastic error components.
The AUC achieved in simulation $0.60 \pm 0.02$  is compatible with the hardware results $0.58 \pm 0.02$ within statistical uncertainties, which, even if not competitive with the full QTTN, still demonstrates a meaningful separation between the LL and TT classes based on the Lund Jet Plane representation.

\section{Conclusions}
\label{sec:conclusions}

The application of quantum machine learning to collider physics requires data representations that are both theoretically robust and compatible with the strict constraints of near-term quantum hardware. In this work, we introduced the Lund Plane to Bloch (LP2B) encoding and processed the resulting representations through a Quantum Tree-Topology Network. By mapping the Cambridge/Aachen declustering history onto a fixed-depth quantum register, our approach explicitly preserves IRC safety. Crucially, the LP2B encoding utilizes a differentiable stereographic projection equipped with learnable stretch parameters, which dynamically optimizes the embedding of the unscaled QCD phase space onto the Bloch sphere. The inherent ``zero-safe'' property of this projection ensures that the topological sparsity of the jet cascade is faithfully retained, handling kinematic dead-ends without injecting spurious rotational noise into the quantum state.

The QTTN demonstrates competitive discriminative power when evaluated across a comprehensive suite of object classification and polarization tagging benchmarks. In the challenging phase space of fully hadronic vector boson scattering and graviton resonance topologies, where kinematics were explicitly decoupled to force reliance on intra-jet radiation patterns, the architecture matches the performance of large-capacity classical deep learning models such as LundNet, and shows a performance advantage against both classical Boosted Decision Trees and constituent-level quantum embeddings (1P1Q). Achieving this with only $\mathcal{O}(10^2)$ trainable parameters, the QTTN effectively pushes the performance-to-cost Pareto front, demonstrating that aligning the entanglement topology with the physical hierarchy of the QCD splitting cascade yields substantial gains in expressive power.

Beyond raw classification accuracy, the proposed quantum architecture addresses two fundamental bottlenecks in contemporary experimental physics: data scarcity and theoretical systematic uncertainties. We demonstrated that the QTTN exhibits resilience in the low-data regime, maintaining robust performance when trained on datasets smaller than those required by classical neural networks, or balancing the benefits of BDTs and MLP architectures. 

Furthermore, cross-evaluations between \texttt{Pythia} and \texttt{Herwig} simulations reveal that the compact parameter space of the QTTN acts as a natural regularizer against overfitting to generator-specific hadronization and parton shower tunings. This reduced susceptibility to non-perturbative mismodeling is particularly valuable for precision jet substructure measurements, where hadronization and parton showering systematics often dominate the overall uncertainty associated to deep taggers.

Finally, the minimal qubit requirement ($N=7$) and shallow circuit depth of the QTTN render it highly viable for current NISQ devices. We successfully validated the core dynamics of the framework on a 3-qubit liquid-NMR SpinQ processor, confirming that the performance of simulated classification logic translates reliably to physical hardware. Looking ahead, the minimal parameter count and compact dimensionality of this tree network architecture open promising avenues for low-latency hardware implementations, such as FPGA-based triggers, and facilitate dedicated applications in data-driven, low-yield searches at the High-Luminosity LHC.

\begin{acknowledgements}
This work was funded by “BOOST – Boosted Object and Oriented-Space Topologies from VBS@HL-LHC - CUP I57G21000110007, finanziato con fondi derivanti dal bando a cascata SPOKE 2 - “Fundamental Research \& Space Economy” nell’ambito del progetto PNRR ICSC, codice CN00000013, CUP I53C21000340006 - Missione 4 – Componente 2 – Investimento 1.4 “Potenziamento strutture di ricerca e creazione di "campioni nazionali di R\&S su alcune Key Enabling Technologies”
The author gratefully acknowledges the UniNuvola Cloud infrastructure of the University of Perugia and INFN for providing the high-performance computing resources used in this study. 
\end{acknowledgements}

\appendix

\section{QTTN ROC Curves}
\label{app:A}
\begin{figure*}[htbp]
     \centering
     \begin{subfigure}[b]{0.45\textwidth}
         \centering
         \includegraphics[width=\textwidth]{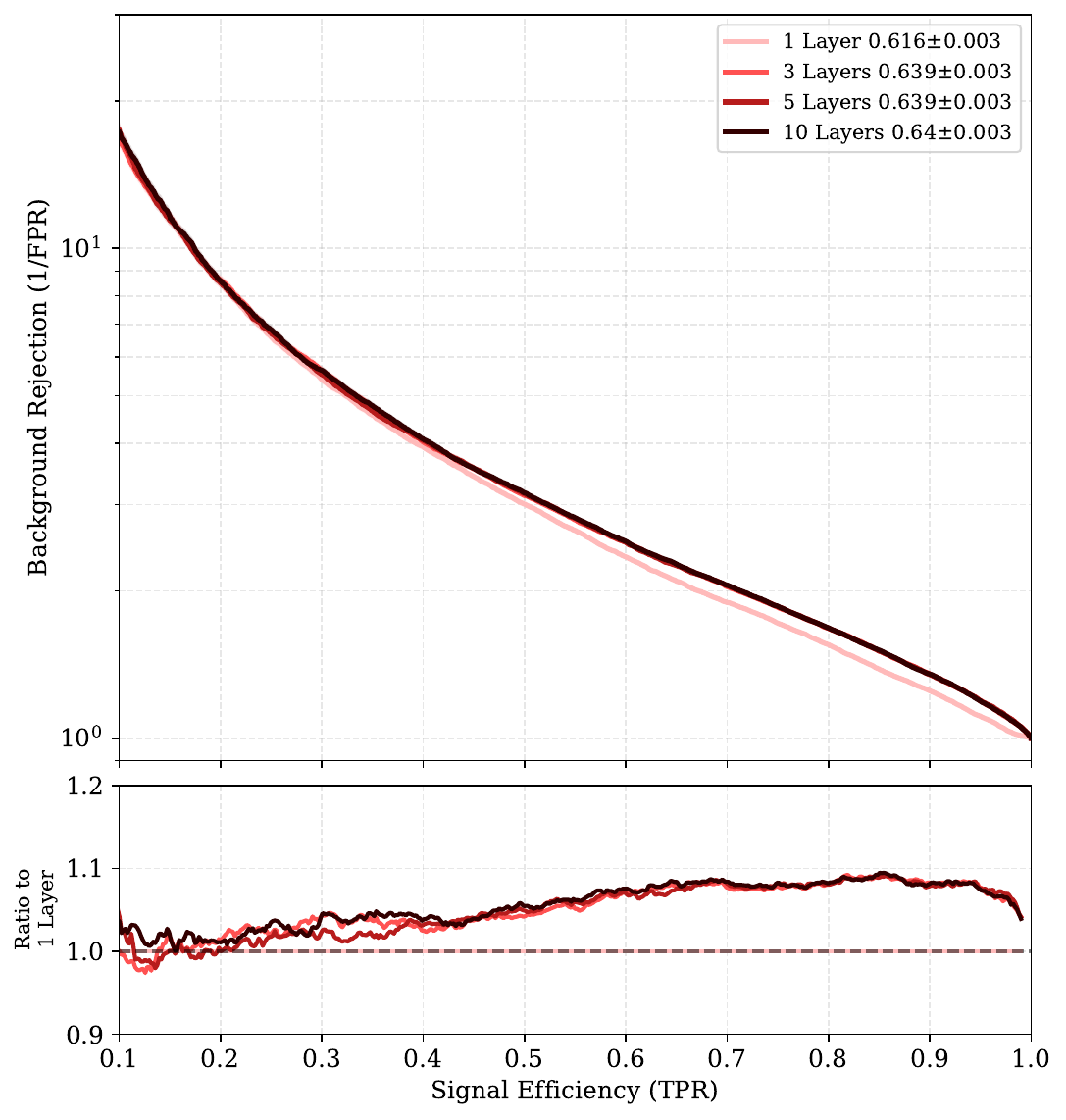}
         \caption{Performance on $W^+W^-$ polarization}
     \end{subfigure}
     \hfill
     \begin{subfigure}[b]{0.45\textwidth}
         \centering

         \includegraphics[width=\textwidth]{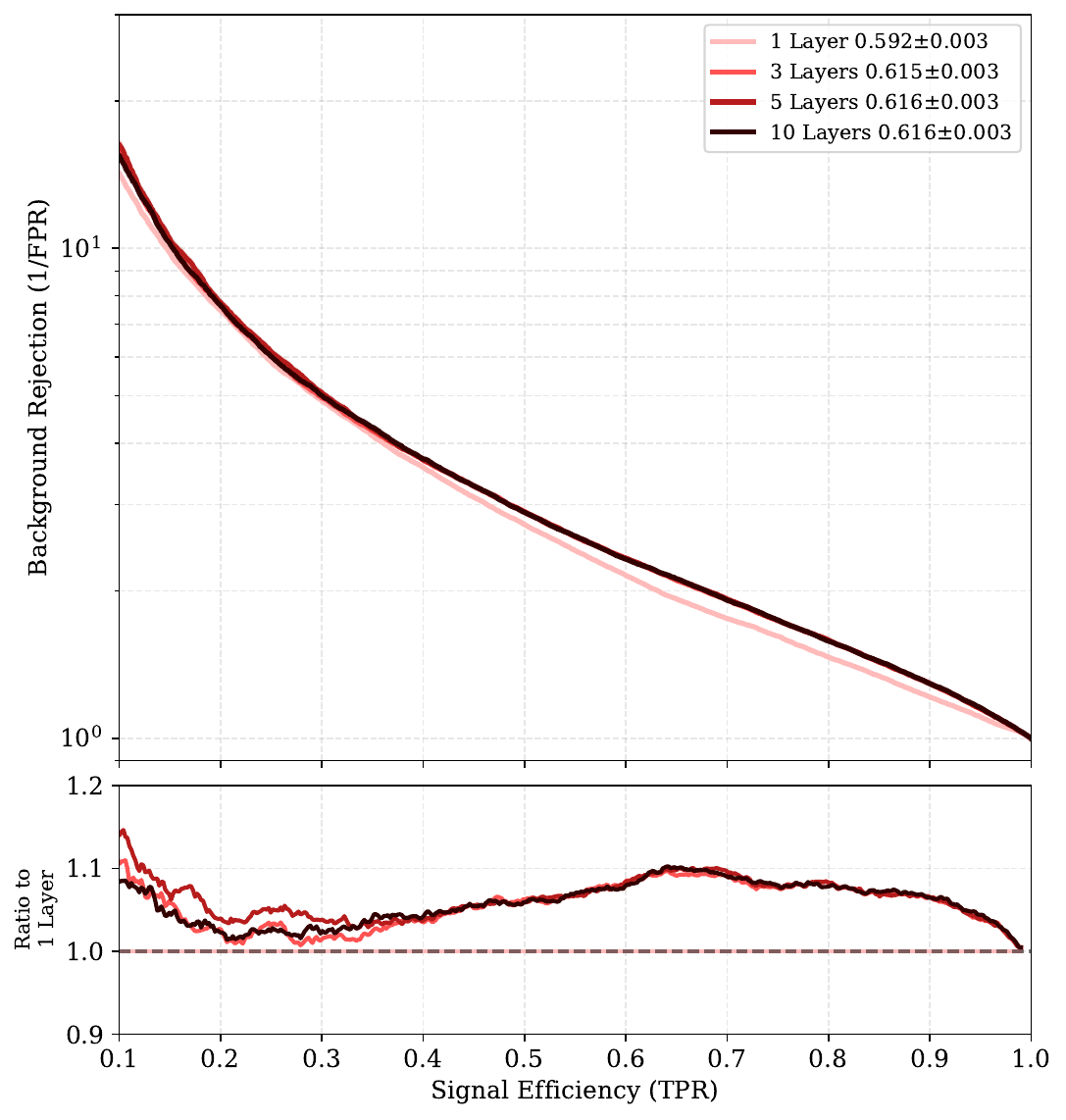}

         \caption{Performance on $G_{\text{bulk}}\to W^+W^-$ polarization}
     \end{subfigure}

     \vspace{10pt} 

     \begin{subfigure}[b]{0.45\textwidth}
         \centering

         \includegraphics[width=\textwidth]{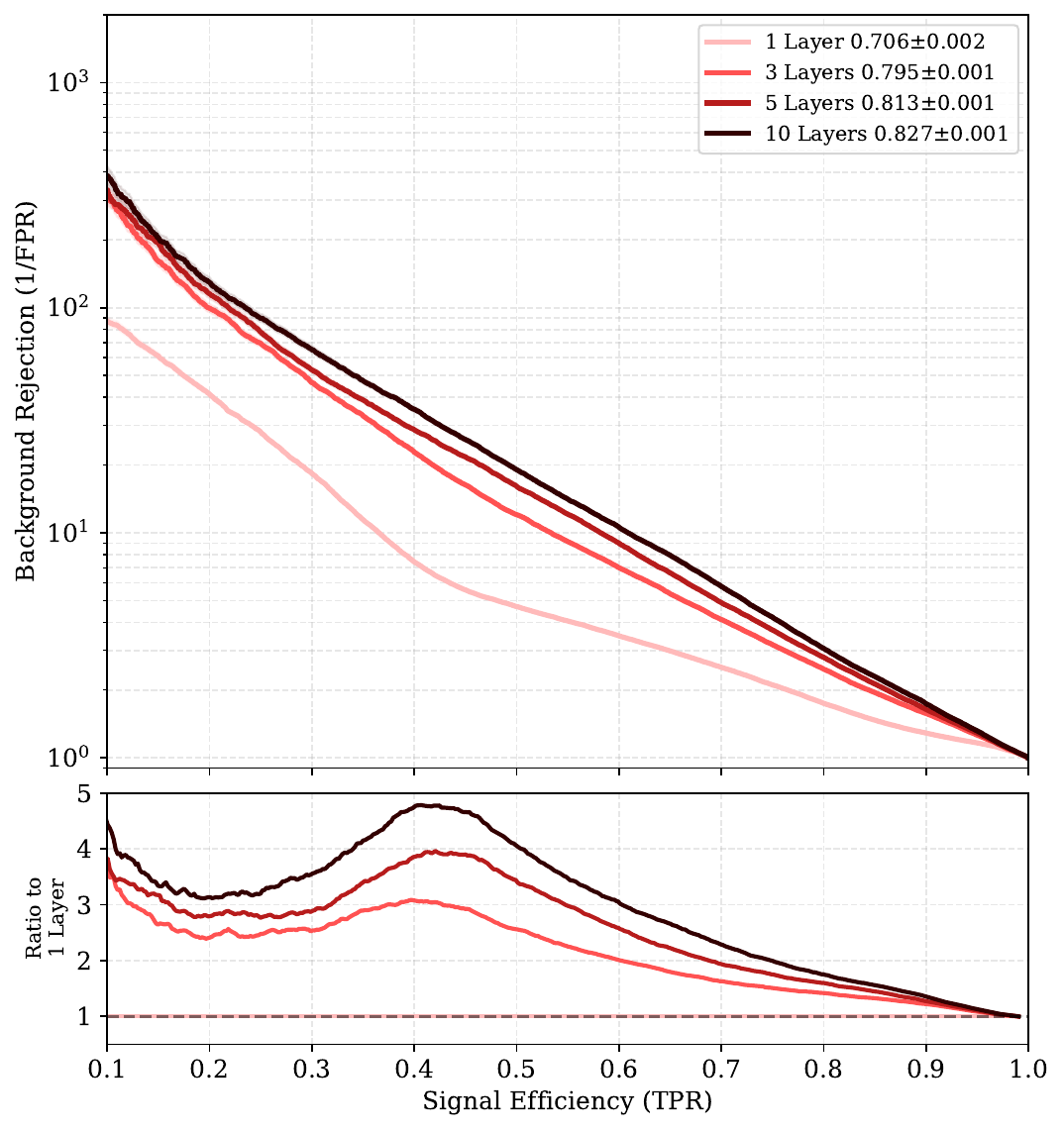}

         \caption{Performance on $W$ tagging}
     \end{subfigure}
     \hfill
     \begin{subfigure}[b]{0.45\textwidth}
         \centering

         \includegraphics[width=\textwidth]{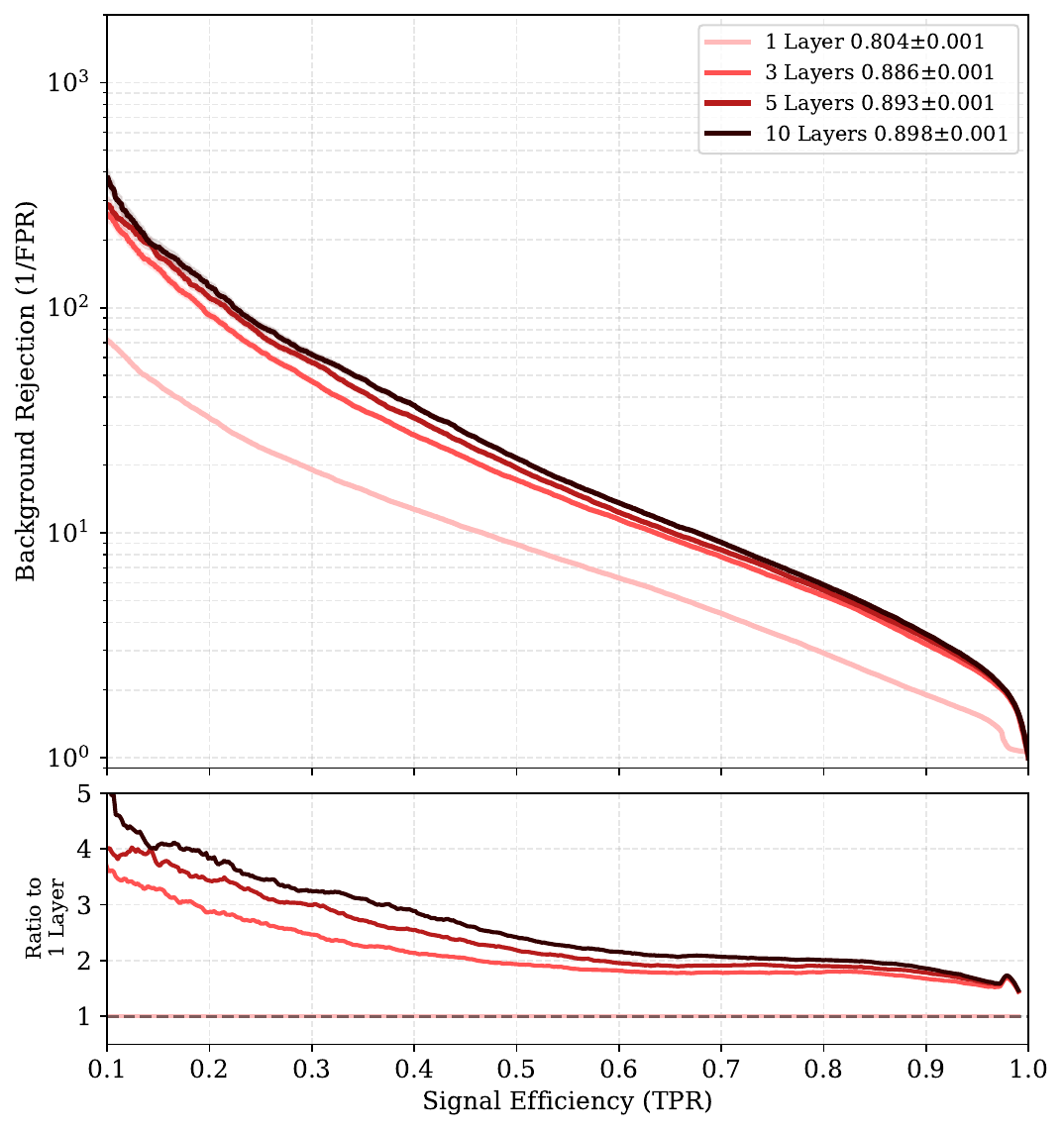}

         \caption{Performance on Top Tagging}
     \end{subfigure}
     
     \caption{Comparison of ROC curves for the QTTN for the different number of layers considered in this work.}
     \label{fig:QTTN_ROC_LAYERS}
\end{figure*}

The performance of the QTTN for the different number of layers $L$ is shown in Figure~\ref{fig:QTTN_ROC_LAYERS} for the different scenarios considered.

\section{Gradient Saliency of Variational Quantum Circuits}
\label{app:B}

To quantify the relative importance of the diverse parameter sets within the QTTN architecture, we perform a first-order saliency analysis. For each parameter $\phi_i$ in the augmented set $\{ \boldsymbol{\lambda}, \boldsymbol{\omega}, \boldsymbol{w}, \boldsymbol{\alpha}, \boldsymbol{\beta}, \boldsymbol{\gamma}, c_w, c_b \}$, the saliency $S_i$ is defined as the mean absolute gradient of the final logit $y_{\text{logit}}$ with respect to $\phi_i$, averaged over a dataset $\mathcal{D}$ of $N$ samples:
\begin{equation}
    S_i = \frac{1}{N} \sum_{j=1}^{N} \left| \frac{\partial y_{\text{logit}}(\boldsymbol{x}_j, \boldsymbol{\Theta}, c_w, c_b)}{\partial \phi_i} \right|
\end{equation}
where $\boldsymbol{x}_j \in \mathcal{D}$ represents the $j$-th input jet. This metric characterizes the sensitivity of the hierarchical circuit and the resulting root-node expectation value $E_0$ to specific weight perturbations. 

\begin{figure*}[htbp]
     \centering
     \begin{subfigure}[b]{0.95\textwidth}
         \centering
         \includegraphics[width=\textwidth]{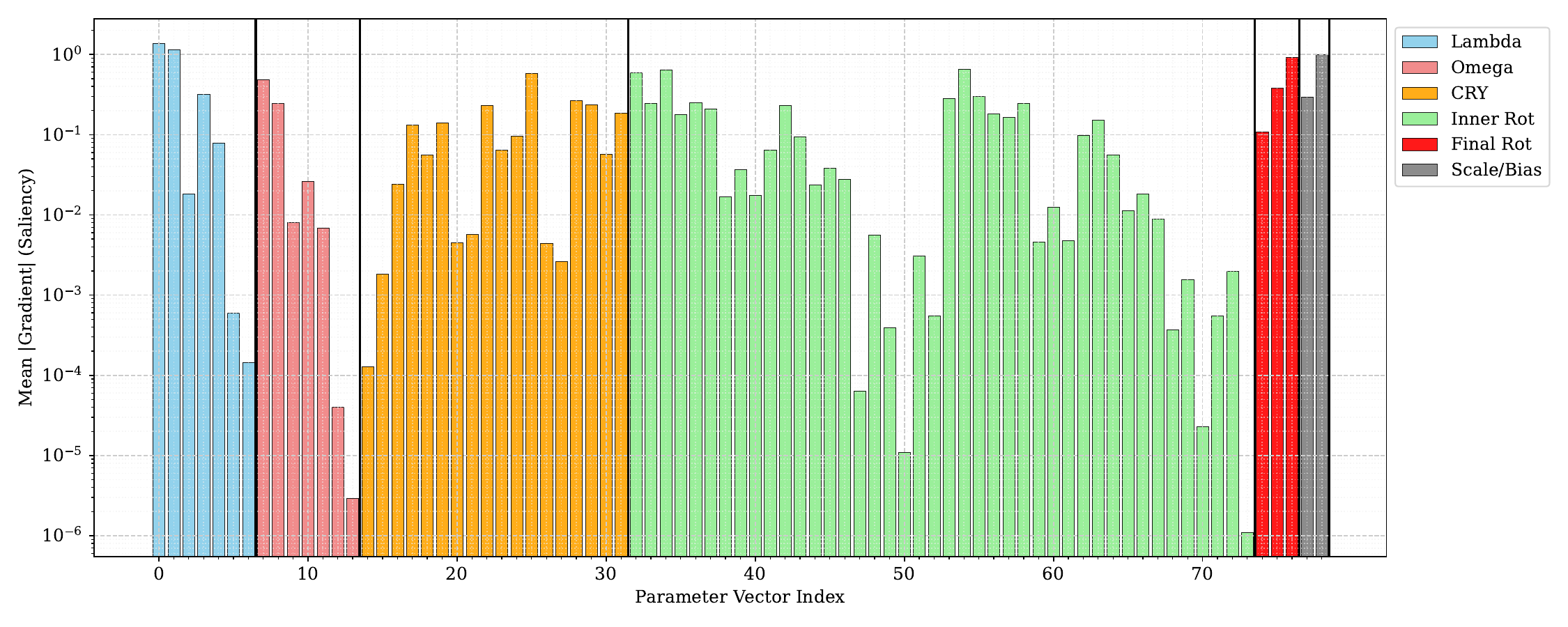}
         \caption{Mean saliency for the QTTN quantum circuit.}
     \end{subfigure}
     \vspace{10pt} 

     \begin{subfigure}[b]{0.95\textwidth}
         \centering

         \includegraphics[width=\textwidth]{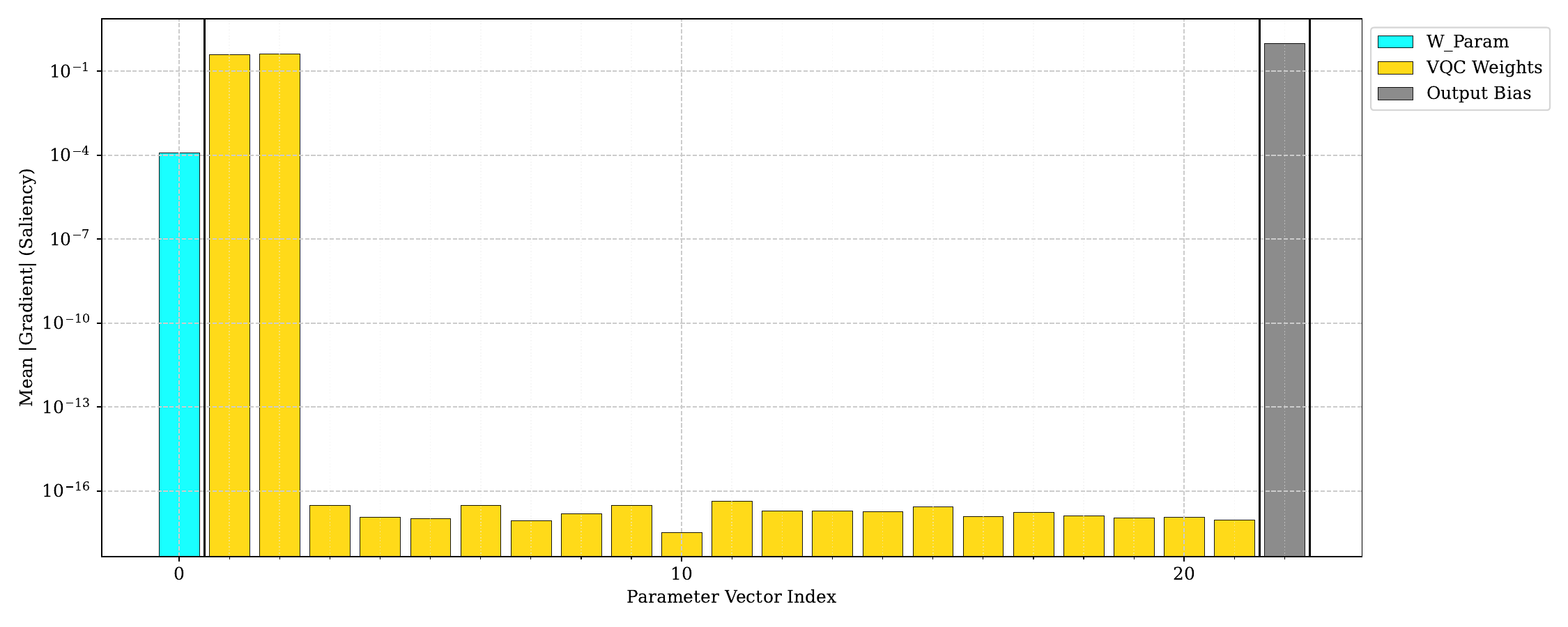}

         \caption{Mean saliency for the 1P1Q quantum circuit.}
     \end{subfigure}

     \caption{Gradient saliency for the parameters of the quantum circuit considered in this work, the QTTN (top) and the 1P1Q (bottom). The color code identifies the different parameter regions in the quantum circuits. For the QTTN, parameters 0-13 are related to the differentiable deformation of the LP2B mapping (blue and red). Parameters 14-31 describe the conditional RY rotations (orange), 32-73 the rotations gates (green), and 74-76 the final rotation before qubit readout. The classical parameters 77-78 (grey) correspond to the scale and bias applied to the quantum measurement. For the 1P1Q the first parameter (cyan) corresponds to the $w$ deformation parameter, 1-21 (yellow) the parametrized rotation, and finally, the last parameter (grey) with the final classical bias applied to the quantum measurement. }
     \label{fig:saliency}
\end{figure*}

The parameter vector is partitioned into functional blocks corresponding to the initial encoding layers $(\boldsymbol{\lambda}, \boldsymbol{\omega})$, the entangling operations $(\boldsymbol{w})$, and the variational rotations $(\boldsymbol{\alpha}, \boldsymbol{\beta}, \boldsymbol{\gamma})$. We also include the classical scalars $c_w$ and $c_b$ to evaluate the influence of the final linear transformation relative to the quantum circuit components. Visualizing $S_i$ on a logarithmic scale, we can identify which layers dominate the gradient signal and determine how effectively the jet information is propagated through the tree structure toward the root qubit $q_0$.
In Figure~\ref{fig:saliency} saliency results are visualized on a logarithmic scale to capture the high dynamic range of sensitivities across different layers, allowing us to identify which components of the QTTN dominate the learning process and which may be susceptible to vanishing gradients or over-parameterization.
By contrast, we observe that the majority of the parameters in the VQC proposed in~\cite{bal2025one} exhibit significantly lower gradient saliency for this task, comparable with the numerical precision used for this study (enabling 64-bit floating-point precision). These parameters correspond to terminal rotations of not-readout qubits.

\bibliographystyle{iopart-num} 

\bibliography{references.bib}        

\end{document}